\documentclass[a4paper,fleqn,usenatbib]{mnras}
\usepackage{newtxtext,newtxmath}

\usepackage[T1]{fontenc}
\usepackage{ae,aecompl}

\usepackage{amsmath}	
\usepackage{amssymb}	
\usepackage{graphicx}	

\usepackage{natbib}
\usepackage{longtable}
\usepackage{color}
\newcommand{\asec}{$^{\prime\prime}$}

\def\r1415{$^{14}$N/$^{15}$N}

\def\15N{$^{15}$NNH$^+$}
\def\N15{N$^{15}$NH$^+$}

\def\HII{H{\sc ii}}

\def\kms{\mbox{km~s$^{-1}$}}
\def\cmc{cm$^{-3}$}
\def\cmq{cm$^{-2}$}

\def\Tex{\mbox{$T_{\rm ex}$}}

\def\Tk{\mbox{$T_{\rm kin}$}}

\def\TMB{\mbox{$T_{\rm MB}$}}

\def\pow#1#2{#1$\times$10$^{#2}$}
\def\kms{km\,s$^{-1}$}

\def\scm{cm$^{-2}$}
\def\ccm{cm$^{-3}$}

\def\hh{H$_2$}

\title[PN in star-forming regions]{Origin of the PN molecule in star-forming regions: the enlarged sample}

\author[Fontani et al.]{F. Fontani$^{1}$,\thanks{E-mail: fontani@arcetri.astro.it}
            V.M. Rivilla$^{1}$,
            F.F.S. van der Tak$^{2,3}$,
            C. Mininni$^{1,4}$,
            M.T. Beltr\'an$^{1}$,
            \newauthor 
            and P. Caselli$^{5}$
          \\
$^{1}$ INAF-Osservatorio Astrofisico di Arcetri, Largo E. Fermi 5, I-50125, Florence, Italy \\
$^{2}$ SRON Netherlands Institute for Space Research, Landleven 12, 9747 AD Groningen, The Netherlands \\
$^{3}$ Kapteyn Astronomical Institute, University of Groningen, The Netherlands \\
$^{4}$ Dipartimento di Fisica e Astronomia, Universit\`a degli Studi di Firenze, I-50125 Florence, Italy \\
$^{5}$ Centre for Astrochemical Studies, Max-Planck-Institute for Extraterrestrial Physics, Giessenbachstrasse 1, 85748 Garching, Germany  \\
          }

\date{Accepted XXX. Received YYY; in original form ZZZ}

\pubyear{2018}

\begin{document}
\label{firstpage}
\pagerange{\pageref{firstpage}--\pageref{lastpage}}
\maketitle

\begin{abstract}
Phosphorus nitride (PN) is the P-bearing species with the highest number of detections in star-forming regions. 
Multi-line studies of the molecule have shown that the excitation temperature of PN is 
usually lower than the gas kinetic temperature, suggesting that PN is likely in conditions of 
sub-thermal excitation. We present an analysis of PN which takes the possible
sub-thermal excitation conditions into account in a sample of 24 massive star-forming regions. 
We observed PN (2--1), (3--2), (4--3), and (6--5) with the IRAM-30m and APEX telescopes
and detected PN lines in 15 of them. Together with 9 similar sources detected in PN in 
previous works, we have analysed the largest sample of star-forming regions to date, made of 33 
sources with 24 detections in total (among which 13 are new detections). Hence, we have increased
the number of star-forming regions detected in PN by more than a factor 2. Our analysis indicates
that the PN lines are indeed sub-thermally excited, but well described by a single excitation temperature.
We have compared line profiles and fractional abundances of PN and SiO, a typical shock tracer, 
and found that almost all objects 
detected in PN have high-velocity SiO wings. Moreover, the SiO and PN abundances with respect to 
H$_2$ are correlated over several orders of magnitude, and uncorrelated with gas temperature. This clearly 
shows that the production of PN is strongly linked to the presence of shocked gas, and rules out alternative 
scenarios based on thermal evaporation from iced grain mantles. 
\end{abstract}


%
\section{Introduction}
\label{intro}

Phosphorus (P) is a crucial element for the development of life on Earth. It is one of the 
key components of the nucleic acids DNA and RNA, phospholipids (the structural components of 
all celullar membranes) and the adenosine triphosphate (ATP) molecule, from which all forms of life 
assume energy (Pasek \& Lauretta~\citeyear{pel}, Mac\'ia et al.~\citeyear{macia}, Pasek et al.~\citeyear{pasek2017}).
Therefore, P plays a key role in three crucial aspects (replication, structure, and energy transfer) of 
living organisms, and could play an important role for life in other planets besides our Earth 
as well (Schaefer \& Fegley~\citeyear{sef}).
Its Solar abundance relative to hydrogen is $3 \times 10^{-7}$ (Asplund et al.~\citeyear{asplund}), i.e. 
about 2--3 orders of magnitude less abundant than the other elements important for life, like oxygen, 
carbon and nitrogen. For this reason, and also because it is likely to be depleted onto dust grains by about 
a factor of $600$ (Turner et al.~\citeyear{turner}, Wakelam \& Herbst~\citeyear{weh}), 
although new measurements indicate a depletion lower than previously thought by a factor of $\sim 100$
(Rivilla et al.~\citeyear{rivilla2016}, Lefloch et al.~\citeyear{lefloch}), its detection in the interstellar medium 
has remained limited so far. 

Among the P-bearing molecules, phosphorus nitride (PN) is the first one detected in the 
interstellar medium toward three high-mass star-forming regions: Orion KL, Sgr B2, and W51, in 
which the measured fractional abundances w.r.t. H$_2$ are $\sim$ (1--4) $\times$ 10$^{-10}$, 
$\sim 10$ times larger than theoretically expected from a pure low-temperature ion-molecule 
chemical network (Turner \& Bally~\citeyear{teb}, Ziurys~\citeyear{ziurys}). Since then, it has been detected in 
high-mass dense cores (Turner et al.~\citeyear{turner}, Fontani et al.~\citeyear{fontani2016}, 
Rivilla et al.~\citeyear{rivilla2016}), as well as in the circumstellar material of carbon- and oxygen-rich 
stars (e.g., Milam et al.~\citeyear{milam}, De Beck et al.~\citeyear{debeck}) 
and in protostellar shocks (Lefloch et al.~\citeyear{lefloch}). Other phosphorus-bearing molecules 
(e.g., CP, HCP, PH$_3$) have been detected in evolved stars (Tenenbaum et al.~\citeyear{tenenbaum}, 
De Beck et al.~\citeyear{debeck}, Ag\'undez et al.~\citeyear{agundez}), but never in dense star-forming cores. 
A considerable step forward was made in the last years thanks to the recent detection of PO in 
high-mass (Rivilla et al.~\citeyear{rivilla2016}) and low-mass (Lefloch et al.~\citeyear{lefloch}) star-forming regions, as well
as in the Galactic Center (Rivilla et al.~\citeyear{rivilla2018}). These new detections have triggered a new interest
in the chemistry of P in the interstellar medium, and more modelling has been carried out to
explain the observational results (e.g.~Rivilla et al.~\citeyear{rivilla2016}, Lefloch et al.~\citeyear{lefloch}, Jim\'enez-Serra
et al.~\citeyear{jimenez2018}), but PN and PO remain the only species detected so far in star-forming regions. 
Due to this lack of observational constraints, how atomic P gets transformed into molecules 
remains poorly understood.

An important result obtained by Mininni et al.~(\citeyear{mininni}) and Rivilla et al.~(\citeyear{rivilla2018}) 
is that PN could likely be originated in shocks. Both studies targeted massive star-forming cores and/or
shocked regions, but located in different galactic environments: either relatively close to the Solar System 
(heliocentric distance $\leq 4.5$~kpc, Mininni et al.~\citeyear{mininni}), or towards the Galactic Centre 
(Rivilla et al.~\citeyear{rivilla2018}). 
Despite the different environmental conditions, both works indicate a correlation between some PN line 
parameters (in particular the line profiles at high velocities), and those of the typical shock tracer SiO. 
Moreover, Mininni et al.~(\citeyear{mininni}) pointed out that the excitation temperature, 
\Tex, of PN, derived by comparing rotational transitions with different energies of the upper level, is 
almost always lower than the gas kinetic temperature, \Tk, estimated from ammonia 
(see Fontani et al.~\citeyear{fontani2011}). A similar result was obtained by Rivilla et al.~(\citeyear{rivilla2018})
towards targets located in the Galactic Centre, in which \Tex\ of PN is $\sim 5$~K while \Tk\
is $\sim$50$-$120 K (e.g.~Guesten et al.~\citeyear{guesten85}, Huettmeister et 
al.~\citeyear{huettemeister93}, Ginsburg et al.~\citeyear{ginsburg16}, Krieger et al.~\citeyear{krieger17}).
This led to the conclusion that the PN lines are sub-thermally 
excited, in agreement with the high critical density of the PN transitions ($\geq 10^{5 - 6}$\cmc), higher than the 
average H$_2$ volume density in the cores analysed by both Mininni et al.~(\citeyear{mininni}) and
Rivilla et al.~(\citeyear{rivilla2018}, for which $n_{\rm H_2}\sim 10^4$ \cmc). This finding
implies that a simple Local Thermodynamic Equilibrium (LTE) analysis is not sufficient to compute
the physical parameters of PN, and calls for complementary non-LTE approaches.

In this paper, we present a multi-line analysis of PN emission taking possible sub-thermal
conditions into account in the largest sample of star-forming cores studied so far. We also extend the 
comparative study between PN and SiO performed by Rivilla et al.~(\citeyear{rivilla2018}) and 
Mininni et al.~(\citeyear{mininni}) to a larger number of sources. 
In Sect.~\ref{obs} we describe the observations analysed in this work, performed with the IRAM-30m
telescope and the Atacama Pathfinder EXperiment (APEX); in Sect.~\ref{res} we present the observational
results, and the two approaches used to analyse the data (LTE and non-LTE); the results are
discussed in Sect.~\ref{discu}, and our conclusions are presented in Sect.~\ref{conc}.

\section{Observations}
\label{obs}

\subsection{Source sample}
\label{sample}

Previous works (Rivilla et al.~\citeyear{rivilla2018} and Mininni et al.~\citeyear{mininni}) suggested 
a correlation between the presence of shocks and/or high velocity material traced by SiO, and the detection 
of PN. However, this tentative conclusion was based on a limited number of sources, and the samples 
were predominantly biased towards sources dominated by shocks. With the aim of avoiding this bias
and having a more robust statistics, we conducted new observations towards a larger sample. 
The new targets include objects with and without clear line wings at high velocity in SiO, 
allowing us to understand if the results obtained in the previous sources, characterised by strong 
SiO wings, applies also to sources with no wings.
We selected targets from a sample of high-mass clumps for which we have IRAM 30m 
observations including SiO (Colzi et al.~\citeyear{colzi2018}, Mininni et al., in prep.)
to allow for a comparison with a typical shock tracer (see e.g.~Downes et al.~\citeyear{downes82},
Ziurys et al.~\citeyear{ziurys89}, Jim\'enez-Serra et al.~\citeyear{jimenez2010}). 
Putting together 24 new observed sources and 9 previously observed ones, we study a 
sample of 33 high-mass clumps, the largest studied so far in star-forming regions. The total sample 
is shown in Table~\ref{tab_sou}.

\begin{table*}
\begin{center}
\caption{List of observed sources and detection summary of the PN rotational lines
(Y = detected; N = undetected; -- = not observed).}
\small
\tabcolsep 0.1cm
\begin{tabular}{ccccccccc}
\hline \hline
Source & R.A. (J2000) & Dec. (J2000) & $V_{\rm LSR}$ & $d$ & \multicolumn{4}{c}{PN} \\
 \cline{6-9}
            &     h:m:s     & ${\circ}$:${\prime}$:${\prime\prime}$ &  km s$^{-1}$ &  kpc & (2--1)$^{(a)}$ & (3--2)$^{(a)}$ & (4--3)$^{(b)}$ & (6--5)$^{(a)}$ \\
            \hline
W3(OH)     & 02:27:04.7 &  +61:52:25.5 & --46.2 & 2.04 & Y$^{(e)}$ & Y & --  & Y \\            
G008.14+0.22      &    18:03:01.3 & --21:48:05.0   &  17.7  & 3.4 & N & N & N  & -- \\
G10.47+0.03       &    18:08:38.0 & --19:51:50.0   &  66.6  & 10.6 & -- & Y & N & N \\
18089-1732M4  &  18:11:54.0 &  --17:29:59.0   &  31.2 & 3.6 & N & N & N & -- \\
G014.33--0.65      &    18:18:54.8 &  --16:47:53.0  &   21.0  & 2.6 & Y & Y  & -- & -- \\
18182--1433M1   &     18:21:09.2 &  --14:31:49.0  &   60.0 & 4.6 & Y & Y & -- & --\\
18272--1217M1  &  18:30:02.9 &  --12:15:17.0  &   32.5 & 2.9 & N & N & -- & -- \\
18310-0825M2  &  18:33:44.0 &  --08:21:20.0  &   80.8 & 5.2 & N & N  & -- & -- \\
G24.78+0.08       &   18:36:12.0  & --07:12:10.0   &  111.0 & 6.3 & -- & Y & N & N \\
G29.96--0.02       &   18:46:03.0  & --02:39:22.0   &  98.0 & 6.2 & -- & Y & -- & N \\
G31.41+0.31       &   18:47:34.0  & --01:12:45.0   &  97.0 & 3.7 & -- & Y & N & Y \\
G34.3+0.2   &    18:53:18.5  & +01:14:58.6   &  60.0 & 3.8 & Y & Y & -- & -- \\
G35.03+0.35        &  18:54:00.0  & +02:01:19.0  &   53.0  & 3.4 & N & N & N & -- \\
G035.20--0.74        &  18:58:13.0 &  +01:40:36.0   &  32.7 & 2.2 & Y$^{(c)}$ & Y & -- & -- \\
G037.55+0.19        &  18:59:11.4  &  +04:12:14.0   &  81.1 & 5.6 & N & Y & N & -- \\
19095+0930   &   19:11:54.0  & +09:35:52.0  &  42.4 & 3.3 & N & N & N & -- \\
G45.07+0.13       &   19:13:22.0  &  +10:50:54.0  &   59.7  & 6.0 & N & N  & -- & -- \\
G45.12+0.13       &   19:13:27.8  &  +10:53:36.7  &   59.5  & 8.3 & N & N  & -- & -- \\
G048.99-0.30      &    19:22:26.3  & +14:06:37.0   &  70.0  & 5.6 & N & Y  & -- & -- \\
W51            &   19:23:43.9  &  +14:30:32.0  &   57.0  & 5.1 & Y$^{(e)}$ & Y$^{(e)}$ & Y & Y \\
20126+4104M1  &  20:14:25.9 & +41:13:34.0  &   --2.7 & 1.7 & N & Y & -- & -- \\
DR21OH       &   20:39:00.4 & +42:22:47.8    & --4.0 & 1.8 & Y & Y & -- & -- \\
NGC7538IRS1  &   23:13:43.3  & +61:28:10.6  &   --53.3 & 2.8 & N & Y & -- & -- \\
NGC7538IRS9   &  23:14:01.8 &  +61:27:20.0   &  --53.3 & 2.8 & N & N & -- & -- \\
\hline
\multicolumn{9}{c}{Previously observed sources} \\
\hline
 AFGL5142-EC  & 05:30:48.7 & +33:47:53 & --3.9 & 1.8 & Y$^{(f)}$ & Y$^{(d)}$ & -- & N$^{(d)}$  \\
 05358-mm3     & 05:39:12.5 &	+35:45:55 & --17.6 & 1.8 & Y$^{(f)}$ & Y$^{(d)}$ & -- & N$^{(d)}$  \\
 AFGL5142-MM  & 05:30:48.0 & +33:47:54 &  --3.9 & 1.8 & Y$^{(f)}$ & Y$^{(d)}$ & -- & N$^{(d)}$  \\
 18089-1732     & 18:11:51.4 & $-$17:31:28 & +32.7 & 3.6 & Y$^{(f)}$ & Y$^{(d)}$ & -- & N$^{(d)}$  \\
 18517+0437  & 18:54:14.2 &	+04:41:41 & +43.7 & 2.9 & Y$^{(f)}$ & Y$^{(d)}$ & -- & Y$^{(d)}$  \\
  G75-core      & 20:21:44.0 & +37:26:38 & +0.2   & 3.8 & N$^{(f)}$ & Y$^{(d)}$ & -- & N$^{(d)}$  \\
  G5.89-0.39   & 18:00:30.5 & $-$24:04:01 & +9.0 & 1.28 & Y$^{(f)}$ & Y$^{(d)}$ & -- & Y$^{(d)}$  \\
  19410+2336    & 19:43:11.4 & +23:44:06 & +22.4 & 2.1 & Y$^{(f)}$ & Y$^{(d)}$ & -- & N$^{(d)}$  \\
  ON1             & 20:10:09.1 & +31:31:36  & +12.0 & 2.5 & Y$^{(f)}$ & Y$^{(d)}$ & -- & Y$^{(d)}$ \\
\hline
\end{tabular}
\label{tab_sou}
\end{center}
$^{(a)}$ observed with the IRAM-30m telescope;
$^{(b)}$ observed with the APEX telescope;
$^{(c)}$ tentative detection at 2.5$\sigma$ rms;
$^{(d)}$ published in Mininni et al.~(\citeyear{mininni});
$^{(e)}$ published in Rivilla et al.~(\citeyear{rivilla2016});
$^{(f)}$ published in Fontani et al.~(\citeyear{fontani2016});
\end{table*}

\subsection{IRAM-30m telescope}
\label{iram}

We observed the PN (2--1) and (3--2) lines at 93.97978 and 140.96775 MHz, respectively
($E_{\rm up}\sim 6.8$ K and $\sim 13.5$ K, respectively), 
using simultaneously band E0 and E1 of the EMIR Single-Side-Band (SSB) receiver of the IRAM-30m Telescope, 
in days June 23--26 and September 26--27, 2017. 
The observations were made in wobbler-switching mode with a wobbler throw of 240\arcsec. 
Pointing was checked almost every hour on nearby quasars, planets, or bright \HII\ regions. 
Focus was checked at the beginning of the observations and after sunset and sunrise. 
The data were calibrated with the chopper wheel technique (see Kutner \& Ulich~\citeyear{kutner}), 
with a calibration uncertainty of about $10\%$. The spectra have been taken in 
main beam temperature units ($T_{\rm MB}$), related to the antenna temperature units 
according to the relation $T^*_{\rm A} = T_{\rm MB}\eta_{\rm MB}$, where 
$\eta_{\rm MB} = B_{\rm eff}/F_{\rm eff}$ is the ratio between the Main Beam efficiency and the 
Forward efficiency of the telescope\footnote{http://www.iram.es/IRAMES/mainWiki/Iram30mEfficiencies}.
The atmospheric conditions were quite stable, with precipitable 
water vapour usually in the range 2 --10~mm, providing system temperatures 
$T_{\rm sys}\sim 150 - 300$ K in band E1 and $T_{\rm sys}\sim 100 - 200$ K in band E0.
The spectra were obtained with the fast Fourier transform spectrometers in mode FTS200, 
providing a channel width of $\sim 200$ kHz (i.e. a resolution in velocity of $\sim 0.6$ and
$\sim 0.42$ \kms\ in band E0 and E1, respectively). 
The PN (6--5) line at 281.9142~GHz ($E_{\rm up}\sim 47.4$ K) was observed in the 
observing run described in Mininni et al.~(\citeyear{mininni}), thus we refer to Sect.~2 of
that paper for any technical aspect of the observations and the data acquisition.

All calibrated spectra were analysed using the GILDAS\footnote{https://www.iram.fr/IRAMFR/GILDAS/} software developed 
at the IRAM and the Observatoire de Grenoble. The spectroscopic parameters used in the derivation of the column 
densities (Sect.~\ref{res_lte}) have been taken from the Cologne Database for Molecular Spectroscopy 
(CDMS; M\"{u}ller et al.~\citeyear{mull05}, Endres et al.~\citeyear{endres}).

\subsection{APEX-12m telescope}
\label{apex}

\begin{figure}
\begin{center}
\includegraphics[width=8cm,angle=0]{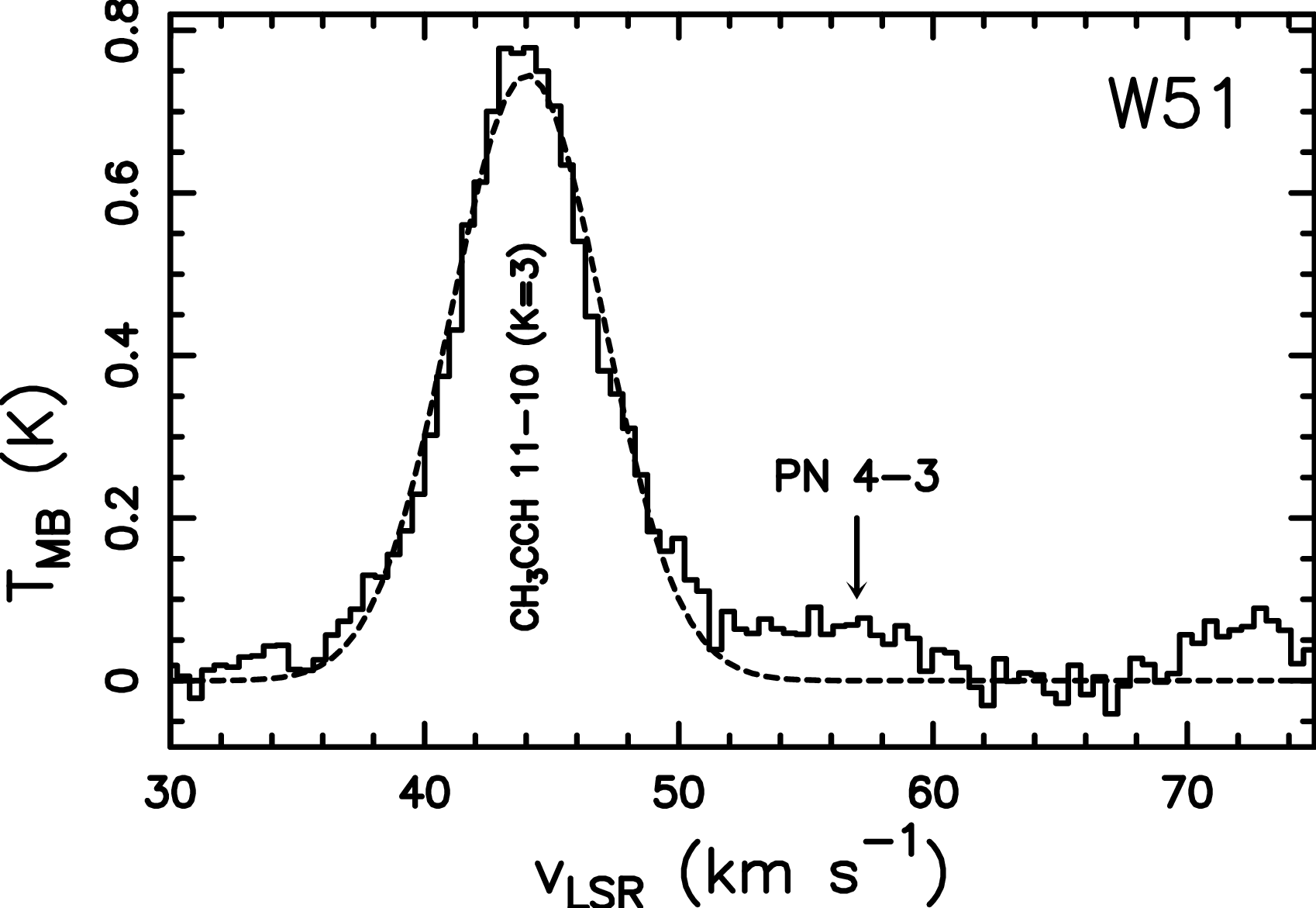}
 \caption{APEX spectrum of PN (4--3) towards W51, the unique target detected
 in this line. The y-axis is in \TMB\ units. The dashed curve indicates the fit to the
 CH$_3$CCH 11--10 (K=3) line, blended with the PN transition indicated by the arrow.}
 \label{spectra43}
\end{center}
\end{figure}

Observations with the Atacama Pathfinder EXperiment (APEX) towards nine sources of the
sample in Table~\ref{tab_sou} were performed at $\sim 1$mm (dual-Side-Band, 2SB, receiver 
SEPIA-B5) to observe the PN (4--3) line at $187.95326$~GHz ($E_{\rm up}\sim 22.6$ K). 
The capabilities of the spectrometer allowed us to simultaneously observe the SiO (4--3) line 
at 173.68831~GHz ($E_{\rm up}\sim 20.8$~K). This line will be used in our comparative 
analysis of PN and SiO in Sect.~\ref{pn-sio}.
The data were acquired with a spectral resolution of $\sim 40$~kHz (i.e. with a resolution
in velocity of $\sim 0.07$~\kms). Observations were obtained in wobbler-switching mode,
with a wobbling amplitude of 100\asec. The amount of precipitable water vapour was around 
1~mm. Pointing was checked almost every hour on nearby sources taken from the APEX
line pointing catalogue. Focus was checked at the beginning of each observing shift on 
planets (Venus, Jupiter or Saturn) or nearby giant stars and/or HII regions.
The APEX raw data are calibrated on-line by the OnlineCalibrator program, which 
writes the calibrated spectra in antenna temperature units into a CLASS-format data file.
PN (4--3) was barely detected only towards the source W51 (see Table~\ref{tab_sou}),
in which the line is blended with a much stronger transition of CH$_3$CCH 11--10 (K=3, 
see Fig.~\ref{spectra43}). Therefore, we have decided not to update the analysis performed 
in Rivilla et al.~(\citeyear{rivilla2016}), based already on transitions (2--1) and (3--2),
because the error associated to the integrated intensity of the (4--3) line is too high, and
hence the analysis already performed in Rivilla et al.~(\citeyear{rivilla2016}) likely would not 
benefit adding this transition.

\begin{figure*}
\begin{center}
\includegraphics[width=14cm,angle=0]{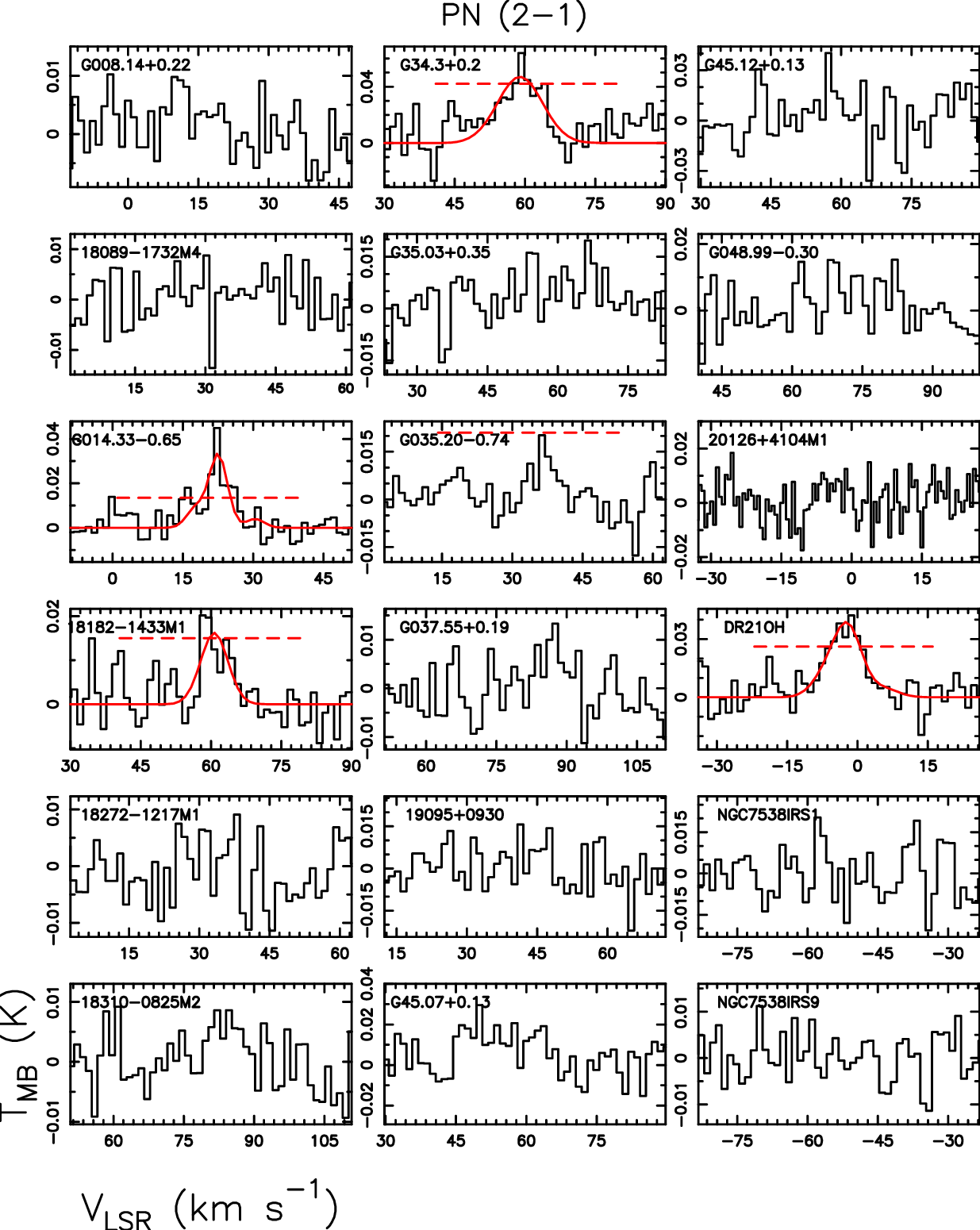}
 \caption{IRAM-30m spectra of PN (2--1) in the 20 sources observed in this line for
 the first time (Table~\ref{tab_sou}).
 The y-axis is in \TMB\ units. The range of velocity on the x-axis corresponds to $(-30,+30)$ \kms\
around the local standard of rest systemic velocity, $V_{\rm LSR}$, given in Table~\ref{tab_sou}.
 The horizontal dashed line indicates the 3$\sigma$ rms level, and it is shown only in the
 detected sources.}
 \label{spectra21}
\end{center}
\end{figure*}

\begin{figure*}
\begin{center}
\includegraphics[width=14cm,angle=0]{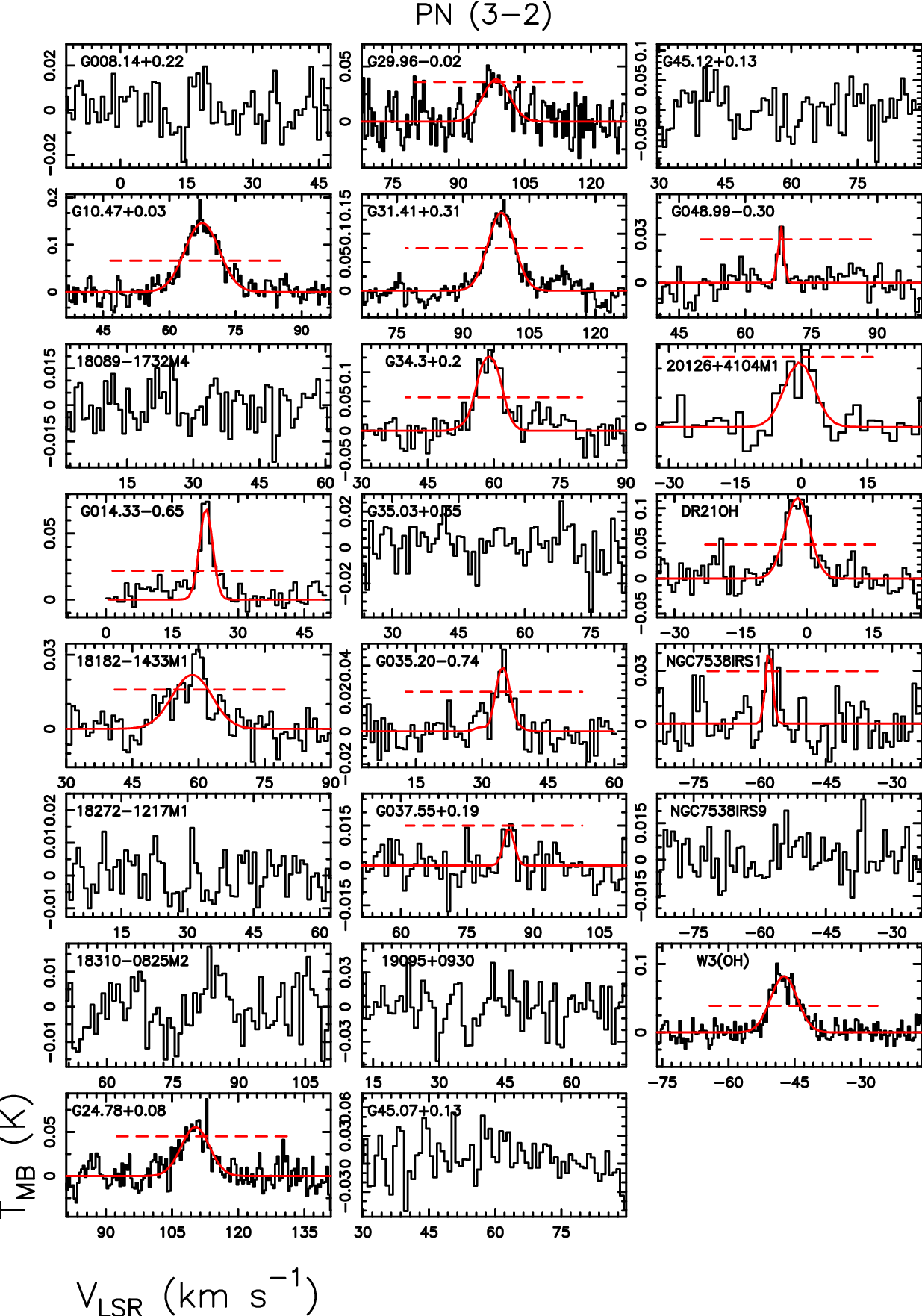}
 \caption{IRAM-30m spectra of PN (3--2) in the 24 sources observed in this line for the
 first time (Table~\ref{tab_sou}).
 The format (units, symbols, labels) are the same as in Fig.~\ref{spectra21}.}
 \label{spectra32}
\end{center}
\end{figure*}

\begin{figure}
\begin{center}
\includegraphics[width=9cm,angle=0]{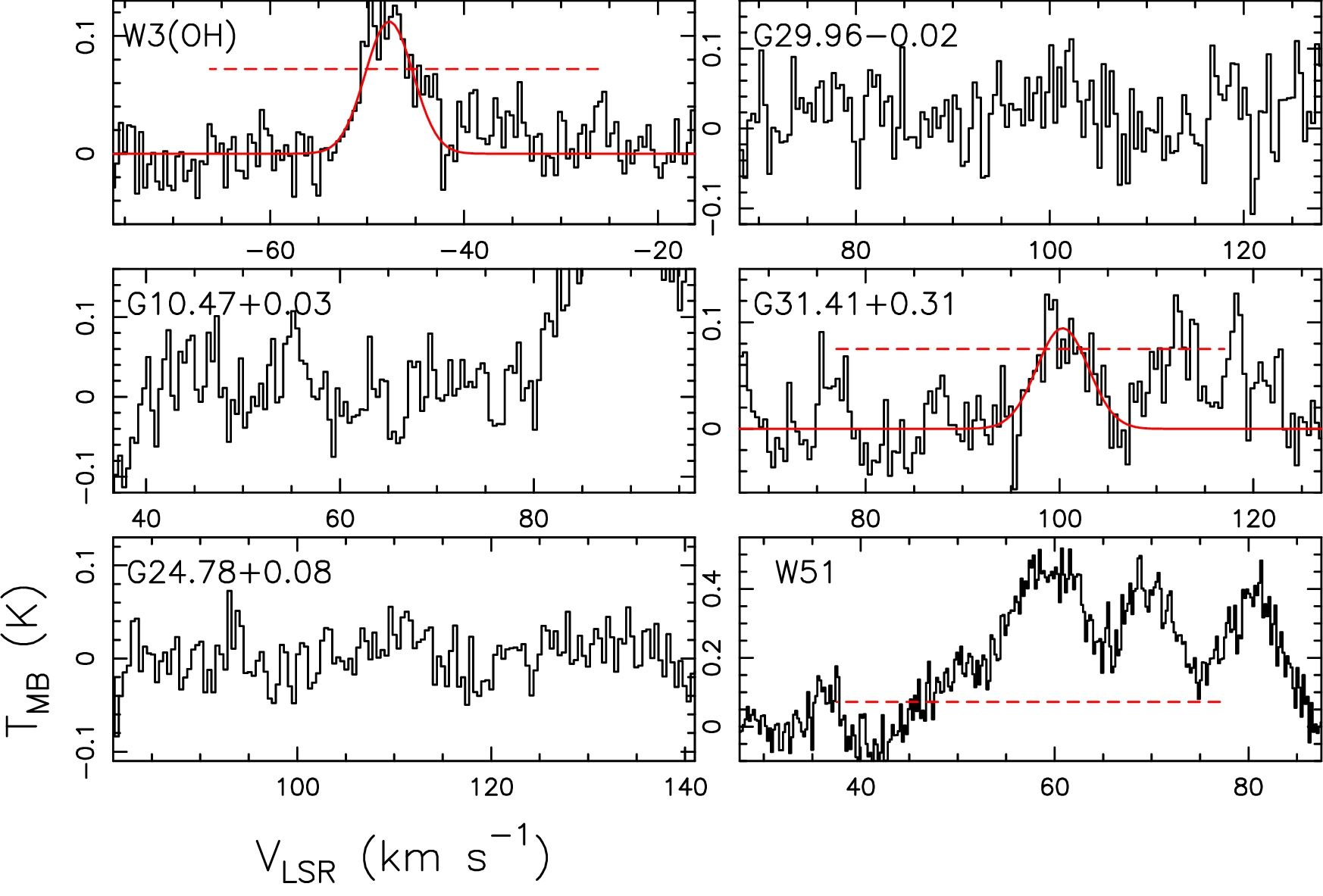}
 \caption{IRAM-30m spectra of PN (6--5) in the 4 sources observed in this line for the first
 time (Table~\ref{tab_sou}).
 The format (units, symbols, labels) are the same as in Fig.~\ref{spectra21}.}
 \label{spectra65}
\end{center}
\end{figure}

\begin{table*}
\begin{center}
\caption{Line parameters derived from the PN spectral analysis described in Sect.~\ref{res}:
integrated line intensity ($\int T_{\rm MB}{\rm d}v$), line width at half maximum ($\Delta V$), and
peak velocity ($V_{\rm peak}$). We also list the $1\sigma$ rms in the spectrum.
We have not included W51, for which the new detections are represented by the (4--3) and (6--5) lines, which 
are both highly blended with strong nearby lines and hence we decided not to update its rotation diagram.}
\label{table:res-1}
\small
\tabcolsep 0.1cm
\begin{tabular}{ccccccccccccccc}
\hline \hline
Source & \multicolumn{4}{c}{PN(2--1)} & & \multicolumn{4}{c}{PN(3--2)} & & \multicolumn{4}{c}{PN(6--5)} \\ 
\cline{2-5} \cline{7-10} \cline{12-15}
             & $\int T_{\rm MB}{\rm d}v$ & $\Delta V$ &  $V_{\rm peak}$ & $1\sigma$ rms & & $\int T_{\rm MB}{\rm dv}$ &  $\Delta V$ &  $V_{\rm peak}$ & $1\sigma$ rms & & $\int T_{\rm MB}{\rm dv}$ & $\Delta V$ &  $V_{\rm peak}$  & $1\sigma$ rms \\ 
            & K km s$^{-1}$                  & km s$^{-1}$      & km s$^{-1}$                  & K     &              & K km s$^{-1}$                  & km s$^{-1}$      & km s$^{-1}$                  & K          &                        & K km s$^{-1}$     & km s$^{-1}$             & km s$^{-1}$                        & K                                 \\
\hline
W3(OH)               &                &               &                 &   &  & 0.61(0.02) & 6.9(0.4) & --47.3(0.2) &  0.013  & & 0.69(0.05) & 5.6(0.5) & --47.8(0.2) & 0.024 \\ 
G008.14+0.22      &                &                 &   & 0.0046  &   &     &                 &   & 0.005  & & & & & -- \\ 
G10.47+0.03       &                &                 &   &  --          &  & 1.43(0.05) & 9.0(0.7) & 67.6(0.3)  & 0.02 & & & & & 0.055  \\ 
18089--1732M4  &          &               &     & 0.0048  &        &              &   &   & 0.008  & & & & & -- \\ 
 G014.33-0.65     &    0.24(0.02)    & 2.5(0.5) & 20.6(0.2) &   0.0048   & & 0.25(0.02)  & 1.5(0.2) & 20.2(0.1) &  0.007 & & & & & -- \\ 
18182--1433M1   &    0.12(0.01)    &  2.2(0.4) & 58.8(0.2) & 0.005  & &  0.25(0.02)  & 5.0(0.6) & 58.3(0.2) & 0.005 & & & & & -- \\ 
18272--1217M1  &          &               &   & 0.005   &          &              &    &  & 0.006  & & & & & -- \\ 
18310--0825M2  &          &               &  &  0.007   &          &              &     &  & 0.010  & & & & & -- \\ 
G24.78+0.08       &                  &        &       &    --         &   & 0.40(0.03)   &   8.9(1.0) & 110.5(0.3) & 0.02 & & & & & 0.039  \\ 
G29.96--0.02       &                  &        &       &    --         &   & 0.29(0.04)   &  6(1) & 98.5(0.5) & 0.01 & & & & & 0.045 \\ 
G31.41+0.31$^{(d)}$       &               &           &  &    --         &  & 1.17(0.05)  & 7.2(0.4) & 98.8(0.2) &  0.011 & & 0.63(0.12) & 5.3(0.7) & 100.0(0.4) & 0.029  \\ 
G34.3+0.2   &    0.47(0.03)    & 3.4(0.6) & 60.3(0.3) & 0.014  & & 0.92(0.06)  &  3.1(0.3) & 60.5(0.1) & 0.019  & & & & & --  \\ 
G35.03+0.35        &                 &             &  &   0.007   &   &       &               &  & 0.010  & & & & & -- \\ 
G035.20--0.74        &  0.08(0.02)     &  1.7(0.5)    &  31.4(0.2) & 0.0085  & &  0.20(0.03)  &   1.9(0.4) & 31.7(0.1)  &  0.009  & & & & & -- \\ 
G037.55+0.19        &                &         &      &   0.0063  & & 0.04(0.01) &  1.3(0.4) & 79.7(0.2) & 0.006  & & & & & --  \\ 
19095+0930   &              &               &    & 0.0085  &          &            &     &  & 0.016   & & & & & --  \\ 
G45.07+0.13       &                 &        &        &   0.015   &           &        &         &  & 0.016   & & & & & --  \\ 
G45.12+0.13       &                 &         &       &   0.016   &           &         &        &  & 0.032   & & & & &  --  \\ 
G048.99--0.30      &                 &         &       &   0.0063  &  & 0.06(0.02)  &  0.5(0.2) & 70.92(0.06) &  0.009   & & & & & -- \\ 
20126+4104M1  &          &               &  &  0.0083  &  & 0.19(0.02)  &   3.7(0.5)  & --3.8(0.3) & 0.008  & & & & & -- \\ 
DR21OH          &   0.36(0.02)   &  2.8(0.4) & --4.4(0.2) &  0.0087  & &  0.80(0.06) & 3.1(0.3) & --4.9(0.1) &  0.016 & & & & & -- \\ 
NGC7538IRS1  &           &               &    & 0.0031    & &  0.12(0.02)  &  1.5(0.4) & --51.3(0.2) & 0.010 & & & & & -- \\ 
NGC7538IRS9   &          &               &    & 0.0062    &  &                   &                &              & 0.009 & & & & & -- \\ 
\hline 
\end{tabular}
\end{center}
\end{table*}

\section{Results}
\label{res}

The spectra of the PN (2--1), (3--2) and (6--5) lines observed for the first time
towards the sources listed in Table~\ref{tab_sou} are shown in Figs.~\ref{spectra21},
~\ref{spectra32} and \ref{spectra65}, respectively. The detection summary is shown in
Table~\ref{tab_sou}: we report 14 new detections in PN (3--2), 5 new detections in
PN (2--1), and 3 new detections in PN (6--5). As noted in Mininni et al.~(\citeyear{mininni}),
the (3--2) transition is always the most intense before correcting for the different beam
size, but they have intensities comparable to the (2--1) ones after correction. The detected lines 
have been analysed following the same approach used in Mininni et al.~(\citeyear{mininni}): 
we have fitted the lines with single Gaussians despite their hyperfine structure. 
The fit results are shown in Table~\ref{table:res-1}.
The method is justified by the fact that the faintest hyperfine components were either 
below the 3$\sigma$ level, or blended among them so that a fit simultaneous to all the hyperfine 
components was not able to give well-constrained parameters. 
In fact, in the few cases in which the hyperfine components were all above the 3$\sigma$ rms 
level, their blending was such that the typical errors associated either with the optical depth 
or with the intrinsic line width (or with both), were comparable to, or larger than, the values themselves. 
In PN (3--2) three sources, G31.41+0.31, 20126+4104M1, and W3(OH) show hints of a double peak 
not due to the hyperfine structure. This could perhaps be due to the presence of multiple 
velocity components inside the telescope beam, although the noise level in the spectra cannot 
allow us to conclude if these double peaks are real. Only higher sensitivity observations will help
us to shed light on this.

The PN total column densities, $N_{\rm{tot}}$, have been derived following two approaches: a 
"classical" analysis which assumes a Boltzmann distribution for the population of the rotational levels with 
a single \Tex\, as in Mininni et al.~(\citeyear{mininni}), and an approach using the radiative transfer 
program RADEX\footnote{\tt https://personal.sron.nl/$\sim$vdtak/radex/index.shtml} 
(van der Tak et al.~\citeyear{vandertak}).
The two methods, and the results obtained from them, are described in the two following sub-sections.

\begin{table}
\begin{center}
\caption{Estimates of \Tex, $N_{\rm tot}$, and X[PN], obtained assuming a Boltzmann population of the levels as described in Sect.~\ref{res_lte}.}
\small
\tabcolsep 0.1cm
\begin{tabular}{cccc}
\hline \hline
Source & \Tex$^{(a)}$ & $N_{\rm tot}$$^{(a)}$ & X[PN] \\ 
            & K       & $\times 10^{12}$ cm$^{-2}$ & $\times 10^{-13}$ \\
\hline
W3(OH)               & $12^{+0.8}_{-0.8}$ & $5.0^{+0.7}_{-0.7}$ &  \\
G008.14+0.22      &    &  $\leq 0.8$ & $\leq 4$ \\ 
G10.47+0.03       &  16   & $10.7\pm 0.4$ & \\ 
18089--1732M4   &    & $\leq 1.3$ & $\leq 6$ \\ 
 G014.33-0.65     &  4$^{+2}_{-1}$ & 5.1$^{+2.3}_{-1.4}$ & $3.5^{1.6}_{1.0}$ \\ 
18182--1433M1   &  8$^{+8}_{-3}$ & 2.2$^{+0.9}_{-0.1}$ & $2.9^{0.1}_{1.2}$ \\ 
18272--1217M1  &    &  $\leq 1.0$ &  \\ 
18310--0825M2  &    & $\leq 1.6$ & $\leq 4.5$ \\ 
G24.78+0.08       &  4  & $3.0\pm 0.2$ & \\ 
G29.96--0.02       &  16  &  $2.2\pm 0.3$ & \\ 
G31.41+0.31$^{(b)}$    & $10.0^{+1.4}_{-1.1}$  & $8.7^{+2.0}_{-1.4}$ & \\ 
G34.3+0.2   & 7$^{+5}_{-2}$ & 8.6$^{+2.9}_{-1}$ &  \\ 
G35.03+0.35     &    & $\leq 1.6$ & $\leq 6$ \\ 
G035.20--0.74     &   $9_{-4}^{(c)}$ & $1.6^{+1.2 (c)}$ & $2^{+1.7 (c)}$ \\ 
G037.55+0.19     & 4  & $1.1\pm 0.3$ & $0.7\pm 0.2$ \\
                     & 16   & $0.27\pm 0.07$ & $0.18\pm0.05$ \\ 
19095+0930   &    & $\leq 2.5$ & $\leq 12$ \\ 
G45.07+0.13    &    & $\leq 2.5$ &  \\ 
G45.12+0.13    &    &  $\leq 5$ & \\ 
G048.99--0.30    & 4  & $1.3\pm 0.3$ & $1.1\pm 0.2$ \\
                           & 16  & $0.39\pm 0.07$ & $0.26\pm0.05$ \\ 
20126+4104M1  & 4  & $1.54\pm 0.2$ &  \\
                           & 16  & $1.29\pm 0.2$ &  \\ 
DR21OH           &    8$^{+7}_{-3}$ & 6.7$^{+2}_{-0.1}$ & \\ 
NGC7538IRS1  & 4  & $0.9\pm 0.2$ & \\
                          & 16   & $0.8\pm 0.1$ \\ 
NGC7538IRS9   &    & $\leq 1.4$ & $\leq 5$ \\ 
\hline 
\end{tabular}
\end{center}
\label{table:res-2}
${(a)}$ when errors are quoted on \Tex, \Tex\ and $N_{\rm tot}$ are derived from rotation diagrams. The method
was applied to sources with two (or more) lines detected. For sources with one line detected only, we assume a fixed 
value for \Tex\ of 4~K or 16~K (see Sect.~\ref{res_lte} for details); \\
$(b)$ rotation diagram made with the (3--2) and (6--5) lines;\\
${(c)}$ upper limit on \Tex\ (and hence lower limit on $N_{\rm tot}$) not defined because the slope 
of the rotation diagram is not physically acceptable.
\end{table}

\subsection{Derivation of $N_{\rm{tot}}$ assuming a Boltzmann distribution}
\label{res_lte}

Assuming that the rotational levels of PN are populated according to a Boltzmann distribution
with a single \Tex, we have used the integrated areas of the lines to build rotational diagrams 
for the sources in which at least two lines have been detected. From these, we have derived the 
excitation temperatures, \Tex, and the total column densities, $N_{\rm tot}$. 
This method assumes optically thin transitions, which is reasonable considering the faintness 
of the lines and hence the expected low abundance of the molecule.
The source angular size, needed to compare transitions observed with different beam sizes, is unknown so far.
Therefore, we have assumed that the PN emission in all the lines fills the smallest beam size, i.e.~9\asec,
corresponding to the beam of the (6--5) line. The integrated line intensities of the (2--1) and (3--2) lines
have then been corrected for beam dilution to determine $N_{\rm tot}$. 
\Tex\ and $N_{\rm tot}$ derived from the rotational diagrams, and all the parameters used to derive them 
with the uncertainties (i.e. integrated areas, integration velocity interval, 1$\sigma$~rms) are shown in 
Table~\ref{table:res-2}. 

For the 7 sources for which we have only one detection in the (3--2) line, the rotation diagram 
cannot be performed and hence the excitation temperature cannot be derived.
Therefore, $N_{\rm{tot}}$ is calculated through \mbox{Eq. (A4)} of Caselli 
et al.~(\citeyear{caselli2002}), assuming \Tex\ equal to the minimum and maximum rotation temperature
derived from the other sources in this work and by Mininni et al.~(\citeyear{mininni}), i.e. 4 and 16~K,
respectively.

The results are shown in Table~\ref{table:res-2}. We have derived 
total column densities of PN in the range $\sim 0.3 - 8.6 \times 10^{12}$~\cmq. The excitation temperatures 
computed from the rotation diagrams vary in the range $\sim 4-9$~K. 
Both results are consistent with the previous study of
Mininni et al.~(\citeyear{mininni}). In particular, the derived excitation temperatures are clearly
lower than the kinetic temperatures measured from ammonia, which are typically in the range
20 -- 40~K (see Fontani et al.~\citeyear{fontani2011},~\citeyear{fontani2015}), 
suggesting a population of the levels not at \Tk. 

Finally, for sources undetected in both lines we have estimated an upper limit on $N$(PN) in this
way: we have used the spectrum with better signal-to-noise in the detected lines, i.e. the (3--2) one,
and computed the area of the Gaussian line having peak equal to the 3$\sigma$ rms level, and
full width at half maximum, FWHM, equal to the average of the FWHM measured in the detected
lines, i.e. $\sim 5$~\kms. The upper limit on $N_{\rm tot}$ was computed from the upper limit on the 
integrated area as for the lines with only one detection, assuming \Tex = 4~K to be conservative.

\subsection{Derivation of $N_{\rm{tot}}$ from the RADEX code}
\label{res_nonlte}

\begin{figure}
\centering
\includegraphics[width=8cm,angle=0]{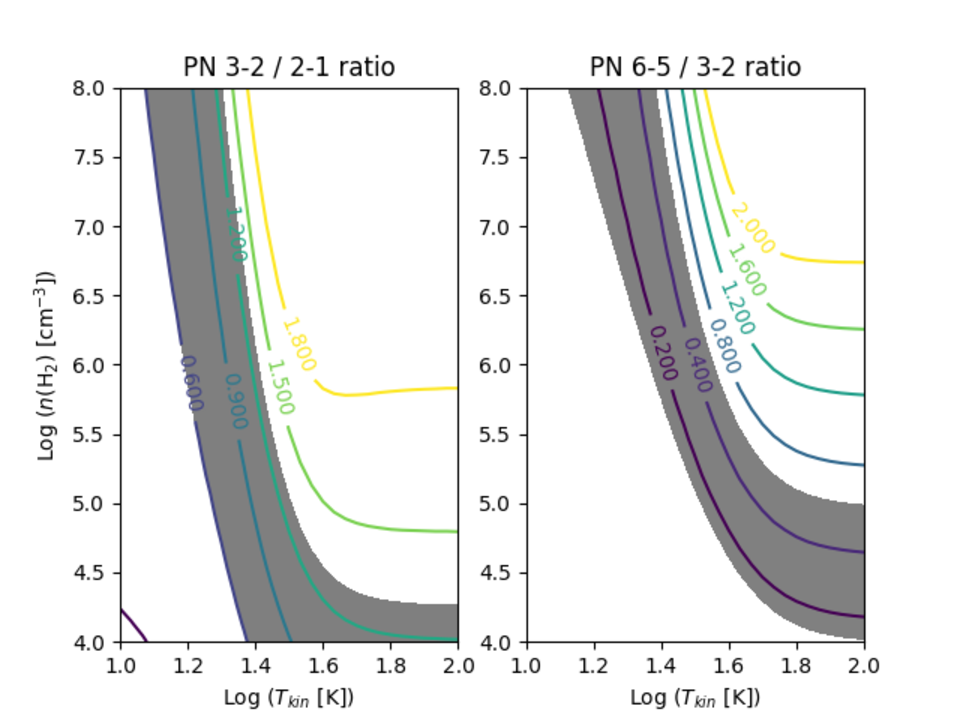}
\caption{Predicted intensity ratios (coloured curves) of the PN 3--2/2--1 (left) and 6--5/3--2 (right) lines, 
calculated with RADEX for PN column densities of $10^{12}$ \cmq.
The gas temperatures range from 10 to 100\,K and the densities from $10^4$ to 10$^8$\,\ccm. 
The observed ranges are indicated by a grey area. As described in Sect.~\ref{res_nonlte}, this
plot is used to derive $n({\rm H_2})$ from the line ratios and \Tk\ given in Table~\ref{table:ratios}.}
\label{f:pn-ratio}
\end{figure}

To consider deviations from a Boltzmann population of the rotational levels, 
the PN total column densities have been also estimated using the non-LTE radiative transfer program 
RADEX\footnote{\tt https://personal.sron.nl/$\sim$vdtak/radex/index.shtml} 
(van der Tak et al.~\citeyear{vandertak}).
This program solves for the radiative and collisional (de)excitation of the molecular energy levels, 
and treats optical depth effects with an escape probability formalism. Collision data for the PN-He system 
(Tobo\l{l}a et al.~\citeyear{tobola}) were scaled by 1.385 to account for \hh\ as dominant collision partner. 

To compute the PN total column densities, the program needs for each source: line widths, line integrated
intensities, \hh\ volume density ($n({\rm H_2})$), and gas kinetic temperature (\Tk). Line widths and integrated
intensities were taken from Table~\ref{table:res-1}. Kinetic temperatures were taken from Fontani et 
al.~(\citeyear{fontani2015}), and are derived from ammonia (1,1) and (2,2) inversion transitions. 
The \hh\ volume densities, $n({\rm H_2})$, were estimated as follows:
Figure~\ref{f:pn-ratio} shows our predicted intensity ratios of the PN 3--2/2--1 and 6--5/3--2 lines, calculated with 
RADEX for kinetic temperatures in the 10--100\,K range and \hh\ densities in the $10^4$--10$^8$\,\ccm\ range. 
The calculations assume a line width of 1\,\kms, a background temperature of 2.73\,K, and a PN column 
density of $10^{12}$\,\scm. Even though the measured line widths are larger than the assumed 1~\kms, the lines are 
expected to be optically thin, and hence the line ratios should not depend neither on column density nor on 
line width. Thus, from the measured line ratios and \Tk, we have estimated $n({\rm H_2})$ from the data plotted in
Figure~\ref{f:pn-ratio} and, from these, the total PN column densities, $N_{\rm tot}$. All parameters used for 
the RADEX analysis, as well as $N_{\rm tot}$, are reported in Table~\ref{table:ratios}. 
Observed intensities are also given in Table~\ref{table:ratios}, and are in between 0.1 and 1.3\,K 
(see also Fontani et al.~\citeyear{fontani2016} and Mininni et al.~\citeyear{mininni}), which translate
into PN column densities of \pow{7}{11}--\pow{3}{13} \cmq\ (see Table~\ref{table:ratios}). 
At these column densities, the optical depths of the PN lines are 0.01--0.1, 
which validate our optically thin assumption for the line ratios above.

The calculated $n({\rm H_2})$ in the PN-emitting gas is generally below $n$(\hh) $\sim$10$^6$\,\ccm, 
i.e. below the critical density of 10$^7$\,\ccm\ where all 3 lines would be in LTE.

\begin{table*}
\begin{center}
\caption{PN line intensity peaks, line ratios, H$_2$ volume density, $n$(\hh), kinetic temperature,
\Tk, and $N_{\rm tot}$ calculated with RADEX (see Sect.~\ref{res_nonlte}) of all massive star-forming 
regions clearly detected in at least two PN lines. All (2--1) and (3--2) peak temperatures have been scaled 
to a source size of 9\arcsec, 
corresponding to the beam of the (6--5) transition, and have uncertainties of the order of the calibration
error (i.e.~$10~\%$ for the (2--1) and (3--2) lines, and $\sim 20\%$ for the (6--5) lines). $N_{\rm tot}$
is also scaled to a source size of 9\arcsec.}
\label{table:ratios}
\tabcolsep 0.2cm
\begin{tabular}{ccccccccc}
\hline \hline
Source &  \multicolumn{3}{c}{$T_{\rm MB}^{\rm peak}$ (mK)} & \multicolumn{2}{c}{peak line ratios} & \Tk\ & $n({\rm H_2})$ & $N_{\rm tot}$ \\
 \cline{2-4}
            & 2--1  & 3--2  & 6--5  &  3--2/2--1 & 6--5/3--2  & K & $\times 10^{5}$ cm$^{-3}$ & $\times 10^{12}$ cm$^{-2}$ \\
\hline
W3OH$^{(a)}$           &   360 & 326 & 112 & 0.9$\pm 0.2$ & 0.3$\pm 0.1$ & 22 & 3.2 & 5$\pm 0.5$ \\
G014.33--0.65           &   290  & 280 &   & 1.0$\pm 0.2$ & & 22 & 4.0 & 1.9$\pm 0.2$ \\
18182--1433M1   &    135 &  100 &   & 0.74$\pm 0.15$ & & 22 & 1.6  & 2.9$\pm 0.3$ \\
G31.41+0.31       &    & 550 & 93  &  & 0.15$\pm 0.05$  & 22 & 7.9 & 7.8$\pm 0.8$ \\
G34.3+0.2          &  450  & 520 &  & 1.2$\pm 0.2$ & & 22 & 5.0 & 6.6$\pm 0.7$ \\
G035.20--0.74        &  180 & 160 &  & 0.9$\pm 0.2$ & & 22 & 3.2 & 1.6$\pm 0.2$ \\
DR21OH       &  360 &  440 &  & 1.2$\pm 0.2$ & & 22 & 5.0 & 5.7$\pm 0.6$  \\
W51            &  1080$^{(a)}$  &  1320$^{(a)}$ & 580  & 1.2$\pm 0.2$ & 0.4$\pm 0.1$ & 22 & 5.0 & \\
AFGL5142--EC &  360$^{(b)}$ &  165$^{(c)}$ & $\leq 66$ & 0.5$\pm 0.1$ & $\leq 0.4$ & 18 & 0.6 & 5.7$\pm 0.6$ \\
05358--m3       &  100$^{(b)}$ &  102$^{(c)}$   & $\leq 48$ & 1.0$\pm 0.2$ & $\leq 0.47$ & 21 & 4.0 & 1$\pm 0.1$ \\
AFGL5142--MM &   402$^{(b)}$ & 365$^{(c)}$ & $\leq 60$ & 0.9$\pm 0.2$ & $\leq 0.16$ & 22 & 3.2 & 2.3$\pm 0.2$ \\
18089--1732  &  180$^{(b)}$ & 240$^{(c)}$ & $\leq 50$ &  1.3$\pm 0.3$ & $\leq 0.21$ & 22 & 6.3 & 2.5$\pm 0.3$ \\
18517+0437  &  180$^{(b)}$ & 176$^{(c)}$ & 108 & 1.0$\pm 0.2$ & 0.6$\pm 0.2$ & 22 & 4.0 & 0.67$\pm 0.07$\\
G5.89--0.39   &  450$^{(b)}$ & 239$^{(c)}$ & 98 &  0.5$\pm 0.1$ & 0.4$\pm 0.1$ & 29 & 3.2 & 38$\pm 4$ \\
19410+2336  &  360$^{(b)}$ & 202$^{(c)}$ & $\leq 45$ & 0.6$\pm 0.1$ & $\leq 0.22$ & 19 & 1.3 & 1.2$\pm 0.1$ \\
ON1              &   180$^{(b)}$ & 134$^{(c)}$ & $\leq 45$ & 0.74$\pm 0.15$ & $\leq 0.34$ & 21 & 1.6 & 2.3$\pm 0.2$ \\
\hline 
\end{tabular}
\end{center}
$^{(a)}$ Rivilla et al.~(\citeyear{rivilla2016});
$^{(b)}$ Fontani et al.~(\citeyear{fontani2016});
$^{(c)}$ Mininni et al.~(\citeyear{mininni});
\end{table*}


%

Comparing Tables~\ref{table:res-2} and \ref{table:ratios}, the $N_{\rm tot}$ estimated with the two methods
appears consistent within a factor $\sim 2$ for all sources except G5.89--0.39, for which the
difference is a factor 5. Therefore, even though the excitation temperatures calculated in Sect.~\ref{res_lte}
are clearly lower than the kinetic temperature, the approximation of a constant \Tex\ seems good to compute
the total column density.


\subsection{SiO emission analysis}
\label{sio}

\begin{figure*}
\centering
\includegraphics[width=14cm,angle=0]{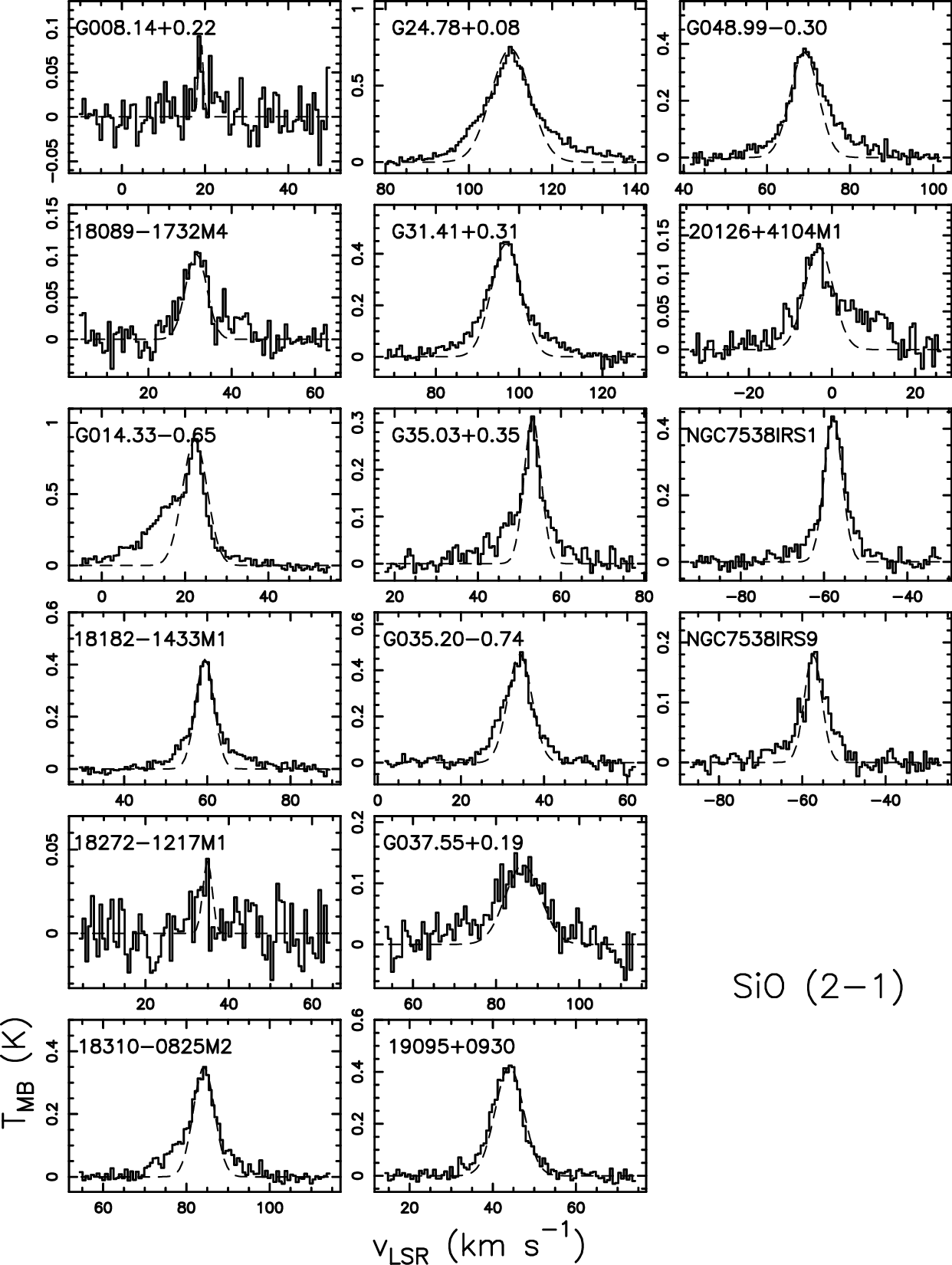}
\caption{Spectra of SiO (2--1) observed with the IRAM-30m telescope. The dashed
curve in each spectrum represents a Gaussian function having maximum and width at half of the
maximum equal to those measured directly from the spectrum (see Table~\ref{table:sio}).}
\label{figure:sio21}
\end{figure*}

\begin{figure}
\centering
\includegraphics[width=8cm,angle=0]{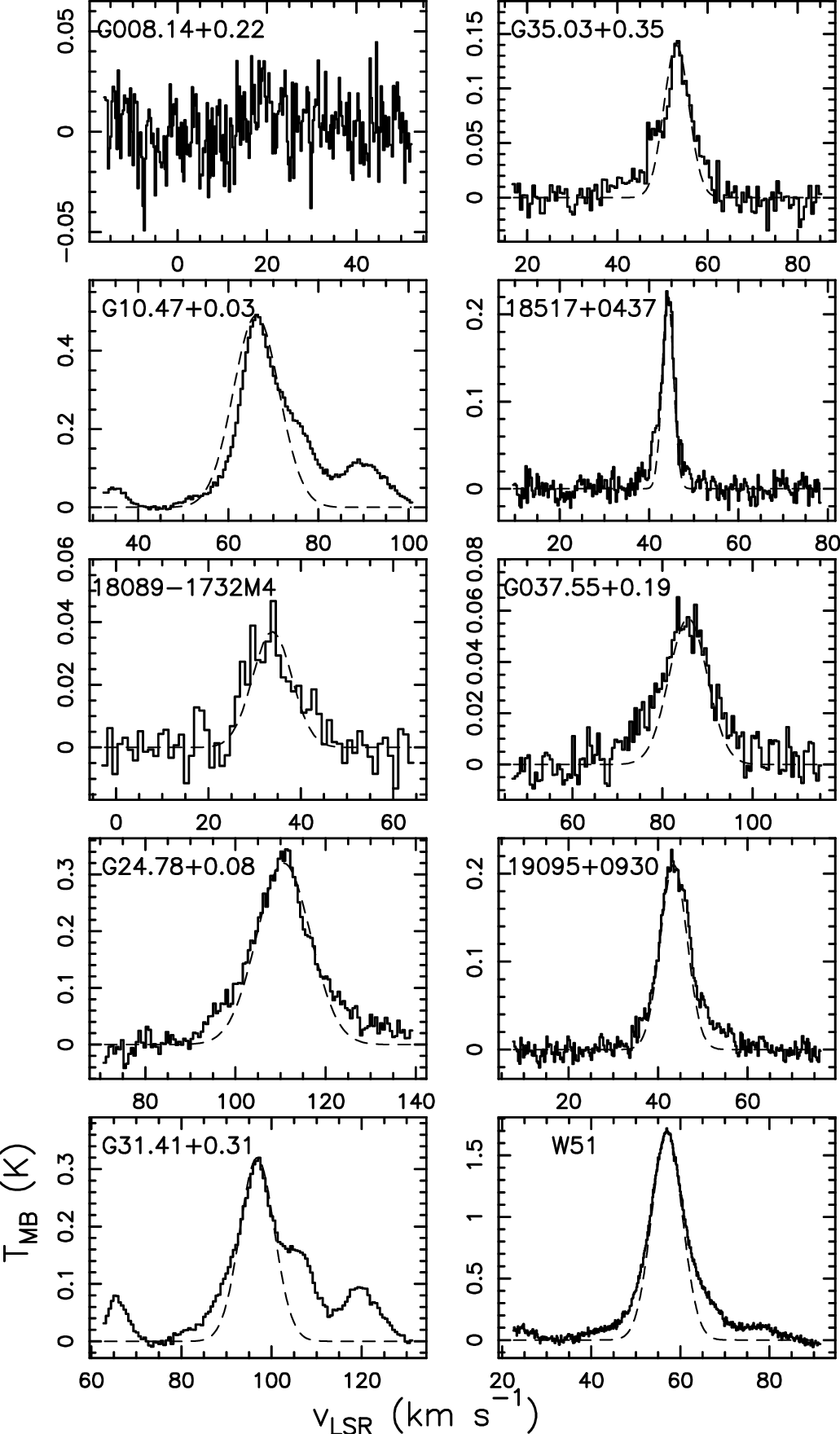}
\caption{Same of Fig.~\ref{figure:sio21} for SiO (4--3) observed with the APEX telescope.
}
\label{figure:sio43}
\end{figure}

In Figs.~\ref{figure:sio21} and \ref{figure:sio43} we show the SiO (2--1) and (4--3) lines 
($E_{\rm up}\sim 6.3$ K and $E_{\rm up}\sim 20.8$ K, respectively) observed towards 
the sources listed in Table~\ref{tab_sou}. The SiO (2--1) lines (rest frequency 86.84696~GHz)
were observed with the IRAM-30m 
telescope during the runs described in Colzi et al.~(\citeyear{colzi2018}). The SiO (4--3) lines
were observed with the APEX telescope in the runs described in Sect.~\ref{apex}. We have
observed in SiO only 19 out of the 24 new sources listed in Table~\ref{tab_sou}. Specifically: SiO (2--1) 
was observed in 16 sources, and detected in all of them, while SiO (4--3) was observed towards 
10 objects (among which 7 targets were observed in the other line) and detected in 9 of them. 
The main line parameters obtained from a Gaussian fit to the lines, i.e. intensity peak, 
velocity at peak position, line width at half maximum, and total integrated intensity, are reported 
in Table~\ref{table:sio}.

Because the main scope of the SiO data in this work is to find the presence of shocked gas,
we have determined if the line profiles show clearly (or likely) high-velocity wings by
overlapping the theoretical profile of a Gaussian line having peak intensity and line width at
half maximum equal to those measured. 
With this approach, we have implicitly assumed that the line profile is not yet affected by shocked 
emission at the half maximum level, which is a reasonable assumption based on the 
spectra in Figs.~\ref{figure:sio21} and \ref{figure:sio43}. By comparing the observed and theoretical 
velocities, we can conclude that all SiO lines have clear or probable high-velocity wings, except 
G008.14+0.22, 18089--1732M4, 18272--1217M1, and 18517+0437 (see Figs.~\ref{figure:sio21} 
and \ref{figure:sio43}). Furthermore, the cases of 18089--1732M4 and 18517+0437 are not obvious
because 18089--1732M4 shows some excess emission at high velocity in the red tale of the (2--1)
line, and towards 18517+0437 Mininni et al.~(\citeyear{mininni}) have found high-velocity wings
in the SiO (2--1) and (5--4) transitions. Hence, only two sources are not associated with clear wings,
and none of them is detected in PN, which indicates that the detection of PN is more likely 
where shocked gas is present.
These results will be discussed in Sect.~\ref{pn-sio-wings}.

\begin{table*}
\begin{center}
\caption{Observed and derived line parameters of SiO (2--1) and (4--3) towards the sample
presented in Table~\ref{tab_sou}. The errors on the best fit parameters are given within parentheses.
In Col.~11 we give the abundance of SiO with respect to H$_2$, X[SiO], derived as explained
in Sect.~\ref{pn-sio-abundances}. Finally, in Col.~12 we report whether the source is detected (Y)
or not (N) in any of the observed PN lines, either in this work or in previous papers.}
\tabcolsep 0.1cm
\begin{tabular}{ccccccccccclc}
\hline \hline
Source  & \multicolumn{4}{c}{SiO (2--1)$^{(a)}$} & & \multicolumn{4}{c}{SiO (4--3)$^{(b)}$} & & & PN? \\
 \cline{2-5}
 \cline{7-10}
            &  $T_{\rm peak}$ & ${\rm V_{\rm peak}}$ & $\Delta V$ & $\int T_{\rm MB}{\rm d}v$ & & $T_{\rm peak}$ & ${\rm V_{\rm peak}}$ & $\Delta V$ & $\int T_{\rm MB}{\rm d}v$ &   & X[SiO]$^{(c)}$ & \\
            &   K                   & km s$^{-1}$    &  km s$^{-1}$       &   K km s$^{-1}$    &  & K  & km s$^{-1}$ & km s$^{-1}$  & K km s$^{-1}$  & wings? & $10^{-12}$ & \\
            \hline
G008.14+0.22      &   0.09(0.02) & 18.9(0.1) &  1.3(0.1) & 0.18(0.04) & & $\leq 0.05$ & -- & -- &  $\leq 0.3$ & N & 0.5$\pm 0.1$ & N \\
G10.47+0.03       &              --              &            --            &          --              &    --   & & 0.710(0.006) & 66.2(0.4) &  8.2(0.4) & 6.34(0.02)  &  Y & & Y \\
18089--1732M4  &  0.27(0.01)  &  32.8(0.2)  & 6.8(0.2) & 2.25(0.5) & &    0.049(0.007)  & 33.8(0.3) &  10(1) & 0.5(0.1) & N? & 1.9$^{+0.1}_{-0.05}$ & N \\
G014.33--0.65      &    0.88(0.03) & 22.3(0.2) & 8.0(0.1) & 9.3(0.2)   & &        --           &          --           &      --     &     -- &       Y        &  13$\pm 2$, 23$\pm 5$ &Y  \\
18182--1433M1   &    0.42(0.01)  & 59.4(0.3) & 5.2(0.2) & 3.25(0.06)  & &          --          &           --          &    --       &    -- &        Y        &  $9\pm 1$, 12$\pm 2$ & Y  \\
18272--1217M1   &   0.04(0.01)  & 34.9(0.3) &  2.6(0.2) & 0.10(0.02) &    &     --            &         --            &    --       &   -- &      N         &   & N  \\
18310--0825M2   &  0.35(0.01)  & 84.2(0.4) & 6.1(0.3) & 3.15(0.05)      &   &        --        &         --            &    --       &   -- &       Y          &   $5.3\pm 0.6$ & N \\
G24.78+0.08       &  0.757(0.007) & 110.0(0.4) & 11.0(0.4) & 11.25(0.04) &   &    0.39(0.02)  & 110.7(0.3) & 11.3(0.3) & 6.57(0.04) & Y  & 14.4$^{+1.0}_{-0.7}$ & Y \\
G31.41+0.31       &   0.44(0.02) & 96.9(0.3)  & 8.4(0.3) & 5.3(0.1)  &  & 0.390(0.006) & 96.9(0.2) &  9.5(0.8) & 4.18(0.02)$^{(d)}$ & Y & $7.0^{+0.5}_{-0.3}$, 9$\pm 1$ & Y \\
G35.03+0.35        &   0.31(0.02) & 53.1(0.3) & 4.4(0.3) & 2.25(0.06) &  &  0.17(0.01)  & 53.1(0.3) & 6.7(0.6) & 1.31(0.03) &  Y & $4.3^{+0.4}_{-0.2}$ & N \\
18517+0437$^{(e)}$    &  --   &   --    &  --   &   --  & &  0.27(0.01)  & 44.2(0.2) & 3.2(0.3) & 1.06(0.02) & N?  & $51\pm 7$, 80$\pm 10$ & Y \\
G035.20--0.74       &   0.47(0.02) & 34.2(0.2)  & 6.8(0.4) &  4.00(0.07)  & &         --      &           --          &      --     &    --  &      Y       &  $11\pm 1$, 13$\pm 2$  & Y \\
G037.55+0.19       &   0.13(0.02)  & 85.5(0.5)  & 12(1) & 1.6(0.1) & &  0.067(0.007) & 85.7(0.2) & 10.5(0.8) & 0.93(0.03) &  Y & $2.0^{+0.2}_{-0.1}$ & Y \\
19095+0930          &  0.42(0.02) & 44.0(0.4)   & 8.0(0.3) & 4.00(0.06)  & & 0.26(0.01)  & 43.0(0.3)  &  7.2(0.2) & 2.38(0.02)  &  Y  & 15$^{+1}_{-0.7}$ & N \\
G048.99--0.30      &  0.37(0.01)  & 69.0(0.3)  & 7.4(0.2) & 4.38(0.06)  & &       --     &       --              &    --       &     -- &    Y       & $6.3\pm 0.7$  & Y   \\
W51           &       --                      &       --               &         --           &    --     & & 2.02(0.05)  & 57.0(0.5)  &  8.0(0.5) & 12.1(0.1) & Y & 910$^{+60}_{-50}$, 1600$\pm 100$ & Y \\
20126+4104M1  & 0.13(0.01)  & --3.1(0.4)  & 8.3(0.4)  & 1.69(0.06) &    &      --      &          --           &      --     &    -- &  Y              &  & Y \\
NGC7538IRS1  &   0.43(0.02) & --57.8(0.3) &  5.1(0.3) & 2.96(0.05)  & &      --       &            --         &      --     &      -- &  Y     &    & Y  \\
NGC7538IRS9   &  0.177(0.007) & --57.4(0.3) & 4.6(0.4) & 1.42(0.03) & &      --       &         --            &      --     &     -- &   Y    & 3.0$\pm 0.4$   & N   \\
\hline 
\end{tabular}
\label{table:sio}
\end{center}
$^{(a)}$ observed with the IRAM-30m telescope; \\
$^{(b)}$ observed with the APEX telescope; \\
$^{(c)}$ for sources with two estimates, the first is based on the SiO total column density computed following the method outlined in Sect.~\ref{res_lte}; the second on the RADEX code, as described in Sect.~\ref{res_nonlte}; \\
$^{(d)}$ line blended with CH$_3$CHO 9(3,6)--8(3,5) at $\sim 173.6824$~GHz; \\
$^{(e)}$ detected by Mininni et al.~(\citeyear{mininni}) in SiO (2--1) and (5--4). We refer to that
work for the SiO (2--1) line parameters. The SiO abundance in Col.~10 is calculated taking also the (4--3) transition 
detected in this work into account.
\end{table*}
 
\subsection{Abundances}
\label{abundances}

We have calculated for both PN and SiO the abundances relative to H$_2$, X[PN] and X[SiO], 
respectively. 
First, we have derived the SiO total column densities. For consistency, we have
followed the same methods used to compute the PN column density, i.e. assuming a Boltzmann 
population of the levels (Sect.~\ref{res_lte}), or from RADEX (Sect.~\ref{res_nonlte}). 
The \Tex\ used for the Boltzmann population is either derived from rotation diagrams, or, for 
sources with one line detected only, fixed to \Tex\ = 8.8~K, that is the average \Tex\ 
derived from the rotation diagrams for the other targets. 

To estimate the SiO column densities with RADEX, we used the line integrated intensities and
line widths provided in Table~\ref{table:sio}, and $n$(\hh) and \Tk\ listed in Table~\ref{table:ratios}. 
We stress that we have all these input parameters for 6 sources only, hence this analysis is
limited to these targets. The collisional coefficients for this analysis are taken from Dayou \& Balan\c{c}a (~\citeyear{deb2006}).

The H$_2$ total column densities, $N({\rm H_2})$, have been derived from a fit to the Spectral Energy 
Distribution between $\sim 160$~$\mu$m and $\sim 850/870$~$\mu$m (Mininni et al., in prep.) on a 
beam of 45\asec. Therefore, the total column densities of both SiO and PN have been rescaled to this 
angular size to estimate X[SiO] and X[PN]. 

The continuum flux density has been modeled as:
\begin{equation}
F_{\lambda} \propto B_{\lambda}(T_{\rm d})(1-e^{-\tau_{\lambda}})
\end{equation}

\noindent
with $\tau_{\lambda}= \kappa_0(\lambda/\lambda_0)^{-\beta} m_{\rm H} \mu_{\rm H_2} {\rm N(H_2)}/\alpha$,
where $\kappa_0=0.8$ cm$^2$ g$^{-1}$ is the dust opacity at $\lambda_0=1.3$~mm, as found 
from the work of Ossenkopf \& Henning~(\citeyear{oeh}), $\beta$ is the dust opacity index,
$m_{\rm H}$ is the mass of the hydrogen atom, $\mu_{\rm H_2}$ is the mean molecular weight, for which we assumed the value of 2.8 
(see Kauffman et al.~\citeyear{kauffmann08}), and $\alpha=100$ is the gas-to-dust mass ratio. 
The range of $\beta$ and $T_{\rm d}$ found are 1.4 -- 2.5 and 23 -- 43~K, respectively.
The fit to the Spectral Energy Distribution (SED) was performed in this way: using 1000 iterations per source, 
at the beginning of each iteration the fluxes were scattered following a Gaussian centred in the mean 
value and with FWHM equal to the error on the measured flux. For each of the 1000 new scattered 
set of values of the fluxes, we performed a fit using a least-square minimising algorithm where the 
three parameters $N({\rm H_2})$, $T_{\rm d}$ and $\beta$ were free to vary. For each of the 3 parameters 
we built a histogram with the 1000 values found, and fitted these histograms with a Gaussian.
The centre of the Gaussian is the best value of the parameter and the FWHM was taken as the error 
on the parameter. A thorough analysis of the continuum data on which our $N({\rm H_2})$ is based
will be provided in a forthcoming paper (Mininni et al., in prep.).
The resulting X[PN] and X[SiO] are listed in Table~\ref{table:res-1} and \ref{table:sio}, respectively.

To check if our estimates are realistic, we have searched in the literature for $N({\rm H_2})$ 
measurements from the (sub-)millimeter continuum, and found results for: 
G008.14+0.22 (Liu et al.~\citeyear{liu2018}), 18182--1433M1 (Beuther et al.~\citeyear{beuther2006}), 
G35.03+0.35 (Pandian et al.~\citeyear{pandian2012}), G37.55+0.19 (Pandian et al.~\citeyear{pandian2012}), 
G048.99--0.30 (Kang et al.~\citeyear{kang2015}), 18089--1732M4 (Beuther et al.~\citeyear{beuther2002}), 
18310-0825M2 (Beuther et al.~\citeyear{beuther2002}). In all these works, $N({\rm H_2})$ is computed
from the dust continuum emission in the (sub-)millimeter, assuming that the emission in this regime is
optically thin. In all but two sources, G35.03+0.35 and G37.55+0.19, the previous estimates are consistent 
with ours considering different assumptions (i.e. slightly different dust opacity or dust temperature). The 
two sources aforementioned, for which our estimates are larger by about an order of magnitude with respect
to the previous estimates, were both observed at 1.1~mm by Pandian et al.~(\citeyear{pandian2012}), and the
authors conclude that, due to the poor angular resolution of their data, their $N$(H$_2$)
could be underestimated by an order of magnitude (see Sect.~4.3 in Pandian et al.~\citeyear{pandian2012}),
which would explain the discrepancy with our estimates.
We thus conclude that the $N({\rm H_2})$ evaluated in this work from the SED are consistent with 
previous estimates obtained from assuming optically thin dust continuum emission
at one wavelength only. The two cases with the highest discrepancy could be due to the
different assumptions in the analysis, but for consistency with the other measurements 
we use our $N({\rm H_2})$ estimates to evaluate the abundances for these two targets as well.

\section{Discussion}
\label{discu}

\subsection{Comparison PN -- SiO}
\label{pn-sio}

As suggested by Mininni et al.~(\citeyear{mininni}) and Rivilla et al.~(\citeyear{rivilla2018}),
PN emission seems to be correlated with the presence of shocked gas as indicated by the
frequent simultaneous presence of high-velocity wings in SiO lines. This suggested a
preferential formation pathway for PN from grain sputtering due to shocks, supported by
the relatively broad width of the lines (FWHM $\geq 3$~\kms). However, this did not seem 
to be ubiquitous, since a few PN lines showed profiles relatively narrow ($\Delta v\>"< 3$~\kms, Fontani et
al.~\citeyear{fontani2016}, Mininni et al.~\citeyear{mininni}). Due to the low statistics, a firm
conclusion on the correlation between the presence of PN and SiO shocked material could
not be given in previous works. In this Section, we compare the PN and SiO lines in our larger 
sample to improve the discussion started in our previous papers.

\subsubsection{Line profiles and high-velocity wings}
\label{pn-sio-wings}

\begin{figure*}
\centering
\includegraphics[width=15cm,angle=0]{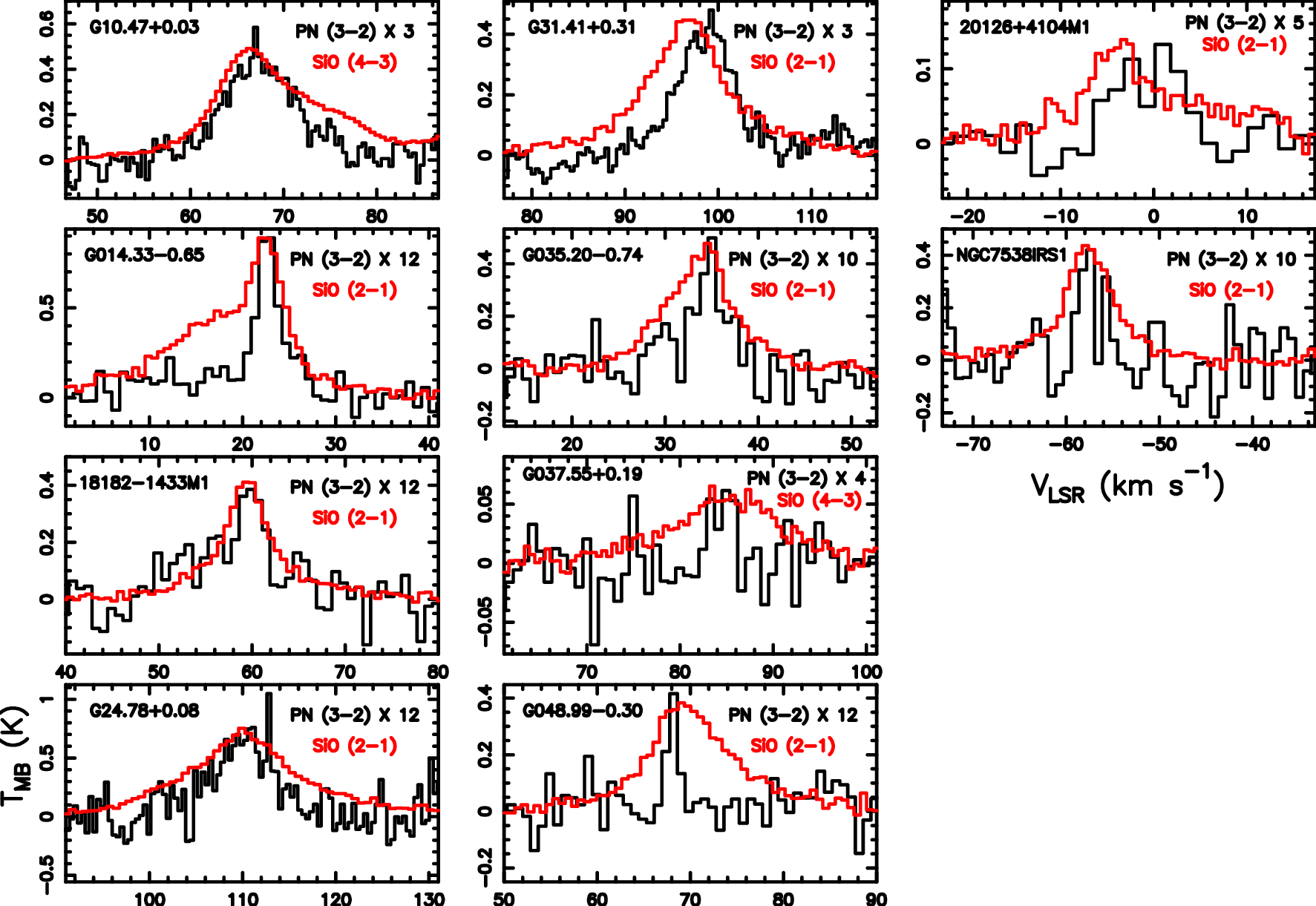}
\caption{Comparison between the profiles of the PN (3--2) line (black) and the SiO (2--1) or
(4--3) lines (in red). The x-axis represents a velocity interval of $(-20,+20)$~\kms\ centred
on the {\rm LSR} velocity of each source, as in Fig.~\ref{spectra21}.}
\label{fig_profile}
\end{figure*}

We compare first the profiles of the PN and SiO lines, to understand if they are similar. Figure~\ref{fig_profile}
shows the comparison between the spectral profiles of PN (3--2) and SiO (2--1) in all sources
in which both lines are observed and detected. For G10.47+0.03, for which SiO (2--1) was not observed, 
we compare the profile of PN (3--2) with that of SiO (4--3). For G037.55+0.19, for which both SiO lines are 
available, we show the most intense one, i.e. the (4--3) transition, but the profile of the (2--1) is
similar (compare Figs.~\ref{figure:sio21} and \ref{figure:sio43}). Inspection of Fig.~\ref{fig_profile} 
suggests that the PN and SiO profiles are almost overlapping in three sources: G10.47+0.03, 18182--1433M1,
and G24.78+0.08. At least six sources show PN emission much narrower than the SiO one, especially G014.33--0.65, 
and G048.99--0.30, in which the high velocity wings clearly detected in SiO are missing in PN.
For the other cases, a discussion is difficult mostly because of the faintness of PN: due to the weak 
PN emission, the high velocity wings could be present below the noise level. 
A peculiar case is G31.41+0.31, in which PN and SiO have a quite similar profile, but the velocity peaks 
are displaced by $\sim 4$~\kms. This could indicate that the bulk of SiO emission is tracing
blue-shifted material more than PN. A similar feature can be noted in 20126+4104M1, in which, however, 
the PN (3--2) line is too faint to derive any firm conclusion. 

In summary, except for G31.41+0.31, the line profiles of SiO and PN could be consistent among
them in all sources, considering the different signal-to-noise ratio at large velocities. This
further supports the presence of PN in sources with SiO wings, already suggested in Sect.~\ref{sio}.

\subsubsection{Abundances}
\label{pn-sio-abundances}

\begin{figure*}
\centering
{\includegraphics[width=8.5cm,angle=0]{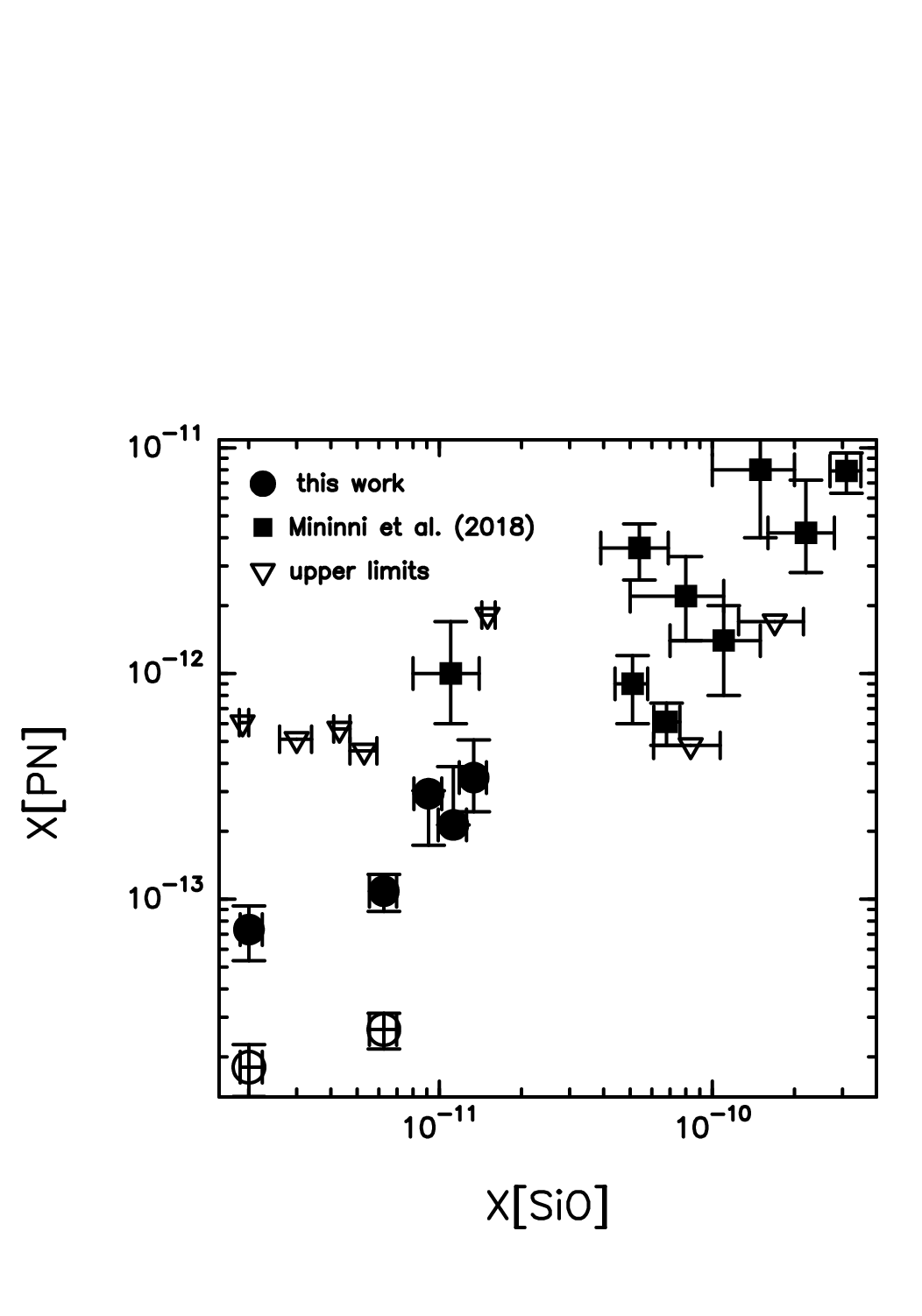}}
{\includegraphics[width=8.5cm,angle=0]{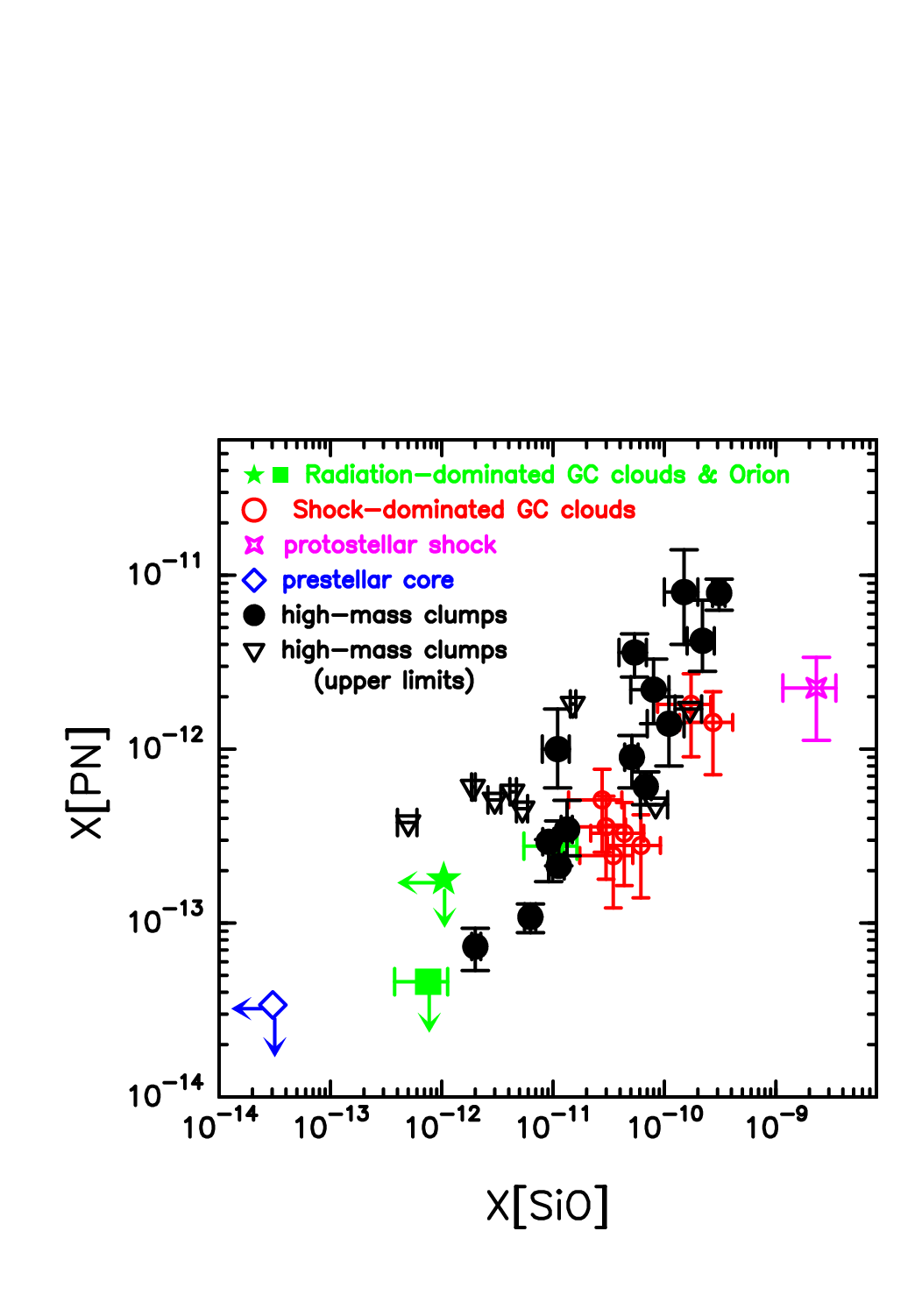}}
\caption{{\it Left:} Comparison between the abundances relative to H$_2$ of PN and SiO, X[PN] 
and X[SiO], respectively. Circles and squares indicate the results obtained in this work and 
by Mininni et al.~(\citeyear{mininni}), respectively. Only the sources in which SiO, PN, and H$_2$ 
column densities could have been derived, or having upper limits on $N$(PN), are shown. 
For G037.55+0.19 and G048.99--0.30, the PN column density has been estimated assuming 
4 and 16~K, hence the abundance has two estimates as well: the open circles indicate X[PN]  
when \Tex = 4~K, and the filled points having the same X[SiO] of the empty ones correspond 
to X[PN] when \Tex = 16~K. The open triangles indicate the $N$(PN) upper limits.
{\it Right:} X[PN] against X[SiO] for the high-mass star-forming clumps illustrated in the left panel (filled circles) 
compared to the measurements of Rivilla et al.~(\citeyear{rivilla2018}) in other regions located in different places
of the Galaxy: shock-dominated Galactic center clouds (red circles) and radiation-dominated Galactic Center 
clouds (green stars), the Orion Bar (green square; Cuadrado, private communication), the L1157-B1 shock 
(magenta open star, Lefloch et al.~\citeyear{lefloch}), and the L1544 pre-stellar core (blue open diamond; 
from the data set from Jim\'enez-Serra et al.~\citeyear{jimenez2016}). The plot is the updated version of Fig.~3 
of Rivilla et al.~(\citeyear{rivilla2018}).}
\label{fig_abundances}
\end{figure*}

The comparison between X[SiO] and X[PN], calculated as explained in Sect.~\ref{abundances}, 
is shown in Fig.~\ref{fig_abundances}: in the left panel, we show the results obtained towards 
high-mass star-forming regions obtained in this work and in Mininni et al.~(\citeyear{mininni}); in the right panel, 
we include all star-forming regions detected so far in PN. Both panels of Fig.~\ref{fig_abundances} show a 
clear positive trend between the two quantities, which confirms and reinforces the relation suggested by 
Mininni et al.~(\citeyear{mininni}) and Rivilla et al.~(\citeyear{rivilla2018}).

The X[SiO] and X[PN] plotted in Fig.~\ref{fig_abundances} were both computed from the 
molecular column densities derived from the Boltzmann population method for several reasons: 
first, it provides the highest number of points; second, it is consistent with the method used in most 
of previous works performed towards other sources; third, two parameters crucial for the RADEX
analysis could not be very accurate (i.e. $n$(\hh) and \Tk). However, as highlighted in Sect.~\ref{res_nonlte}, the 
column densities estimated with the two methods are consistent within a factor 2-3, except for two
objects, and should not change overall the trend, which is based on several orders of magnitude.

The data gathered in this work expand the relation between X[PN] and X[SiO] for high-mass star-forming
regions to an interval of low abundances of PN down to $10^{-13}$, not reached by Mininni et al.~(\citeyear{mininni}).
Moreover, the relation shown in the right panel of Fig.~\ref{fig_abundances} spans more than
two orders of magnitude in X[PN] and more than four orders of magnitude in X[SiO], indicating a real robust
correlation independent on the initial abundance of the host environment.

As explained in Sect.~\ref{res_lte}, for G037.55+0.19 and G048.99--0.30 we estimated $N_{\rm tot}$ 
using two fixed excitation temperatures of 4 and 16~K, so that in Fig.~\ref{fig_abundances} we 
show X[PN] estimated assuming both temperatures. The Pearson's $\rho$ correlation coefficient 
between X[PN] and X[SiO] is about 0.6 considering either of the two estimates. The $p-$value 
is $\sim 0.02$, i.e. below 0.05, and the correlation is thus significant. 
The trend remains positive when including in the comparison other dense cores located in different 
places of the Galaxy: as shown in the right panel of Fig.~\ref{fig_abundances} (which is an updated
version of Fig.~3 of Rivilla et al.~\citeyear{rivilla2018}), the positive trend is still apparent,
indicating that it is independent on the type of source analysed. 

\subsection{Relation with hot core temperature}
\label{pn-T}

One of the possible formation pathways of PN initially proposed by Charnley \& Millar~(\citeyear{cem})
was high-temperature gas-phase chemistry after the thermal desorption of PH$_3$ from ices. In this 
scenario, PH$_3$ is the most abundant species produced on dust grain mantles by hydrogenation of
atomic P which, upon desorption at temperatures of about $\sim 90$~K (Turner et al.~\citeyear{turner}), 
very quickly produces other P-bearing species, with PN and PO being the most efficiently formed ones. 
Hence, this scenario is expected to be particularly relevant in hot cores.
Jim\'enez-Serra et al.~(\citeyear{jimenez2018}) predict a similar formation pathway from neutral-neutral
reactions in warm gas. Even though the analysis provided so far indicates clearly that the PN
emission arises in shocked gas, our single-dish observations encompass an angular region of 20--30\asec,
i.e. much larger than the size of the shocked spots, expected to be more compact than this at the
distance of the sources (i.e. of a few arcseconds). Hence, if a significant part of the total emission arises 
from hot cores inside the telescope beam, and is more efficient with increasing temperature,
we expect a correlation also between X[PN] and the temperature 
measured from CH$_3$CN, T$_{\rm CH_3CN}$, a well-known temperature tracer of hot molecular cores 
(e.g.~Remijan et al.~\citeyear{remijan}, Cesaroni et al.~\citeyear{cesaroni}). 
T$_{\rm CH_3CN}$ was measured in Colzi et al.~(\citeyear{colzi2018}).

\begin{figure}
\centering
{\includegraphics[width=8.5cm,angle=0]{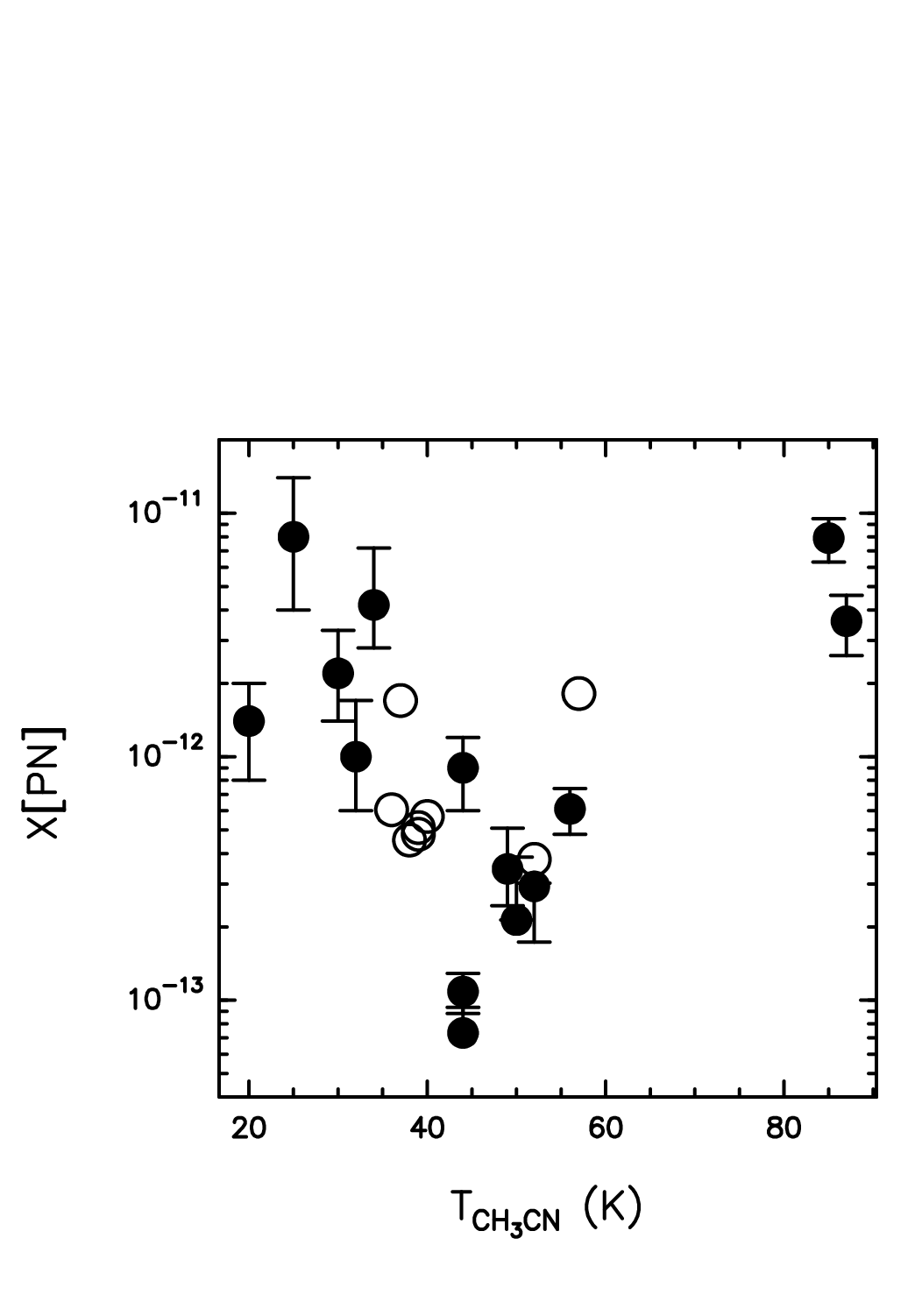}}
\caption{PN abundance against gas temperature estimated from CH$_3$CN 
(Colzi et al.~\citeyear{colzi2018}). Empty symbols correspond to X[PN]
upper limits. }
\label{fig_T}
\end{figure}

As it can be noted in Fig.~\ref{fig_T}, we do not find any correlation between 
these two parameters (correlation coefficient $\sim 0.2$ and $p-$value $\sim 0.5$, i.e. the correlation is not 
significant). This suggests that gas-phase chemistry does not play a significant role in the production 
of PN. Inspection of Fig.~\ref{fig_T} rather suggests that, if one 
excludes the two hottest cores, the trend between X[PN] and CH$_3$CN temperature could appear 
negative instead, but the correlation coefficient is $\sim -0.5$ with $p-$value of $\sim 0.1$, hence it is
not statistically significant either. All this is clearly at odds with a scenario in which PN 
is the quick product of PH$_3$ evaporated from ices at high temperatures, or from other high-temperature 
neutral-neutral reactions. This discrepancy is instead consistent with the fact that PN
and CH$_3$CN do not trace the same material. In fact, CH$_3$CN is typically found in hot cores, i.e.
in close proximity of young stellar objects, while PN (and PO) appear to trace more extended regions 
(see Rivilla et al.~\citeyear{rivilla2016}).


\section{Conclusions}
\label{conc}

We have analysed observations of PN, performed with the IRAM-30m and APEX telescopes, 
in a statistically significant sample of high-mass star-forming clumps. 
The sample includes 11 sources already detected in PN (Fontani et al.~\citeyear{fontani2011},
Rivilla et al.~\citeyear{rivilla2016}, Mininni et al.~\citeyear{mininni}), and 22 new targets never observed
in PN so far. Because previous evidence indicate possible {\rm sub-thermal excitation}
conditions of PN in our targets due to the high ($\sim 10^{7}$\cmc) critical density of its
transitions, we have calculated the PN total column densities following a "classical"
approach assuming a Boltzmann distribution, and a non-LTE approach based on the radiative transfer 
program RADEX. The first approach provides $N_{\rm tot}\sim 0.3 - 8.6 \times 10^{12}$\cmq, and 
excitation temperatures of $\sim 4-9$~K. The RADEX analysis provides $N_{\rm tot}\sim 0.7 - 37 \times 10^{12}$\cmq,
i.e. overall larger column densities, but for all sources the difference is within a factor 2.5, except for one. 
This indicates that the lines are sub-thermally excited but the population of the levels is well
described by a single \Tex, lower than \Tk. The PN abundances are in the range $10^{-13}-10^{-12}$.
We have also detected SiO (2--1) and (4--3) in most of the sources observed in PN.
The main goal of the comparison between PN and SiO is to test that the presence of 
PN is tightly associated with shocked gas, as tentatively proposed on previous works based on a limited 
statistics or focussed on specific regions of the Galaxy. We have found that PN is never detected in 
sources without high-velocity wings in SiO lines, and the SiO and PN line profiles are similar in many
sources. We have also found a positive correlation between X[PN] and X[SiO] which, together with 
the strong link between detection of PN and presence of high velocity wings in SiO, demonstrates, based 
on an unprecedented robust statistics, that SiO and PN have a common origin.

{\it Acknowledgments.}  We thank the IRAM-30m staff for the precious help during the observations,
and the APEX staff for the observations. We also acknowledge L. Colzi for her help during
the observations performed in remote. This work has been partially supported by the project 
PRIN-INAF 2016 The Cradle of Life - GENESIS-SKA (General Conditions in Early 
Planetary Systems for the rise of life with SKA).
V.M.R. acknowledges the financial support received from the European Union's Horizon 2020 research 
and innovation programme under the Marie Sklodowska-Curie grant agreement No 664931.
C.M. acknowledges support from the Italian Ministero dell'Istruzione, Universit\`a e Ricerca 
through the grant Progetti Premiali 2012 - iALMA (CUP C52I13000140001). 
\let\oldbibliography\thebibliography
\renewcommand{\thebibliography}[1]{\oldbibliography{#1}
\setlength{\itemsep}{-1pt}} 

{}

\end{document}